\documentclass[%
 twocolumn,
 superscriptaddress,
 showpacs,
 amsmath,amssymb,
 prl
]{revtex4-2}
\usepackage{setspace}
\usepackage{verbatim}
\usepackage{siunitx}
\usepackage{booktabs}
\usepackage{tabularx}
\usepackage{multirow}
\usepackage{makecell}
\usepackage{graphicx}
\usepackage{dblfloatfix}
\usepackage{dcolumn}
\usepackage{bm}
\usepackage{bbm}
\usepackage{bbold}
\usepackage{upgreek}
\usepackage{physics}
\usepackage{enumerate}
\usepackage[dvipsnames]{xcolor}
\definecolor{darkblue}{HTML}{023e8a}
\definecolor{darkred}{HTML}{780000}
\definecolor{darkgreen}{HTML}{005f73}
\definecolor{darkcyan}{HTML}{2a9d8f}
\usepackage[colorlinks=true,
	        urlcolor=darkred,
	        citecolor=darkblue,
	        linkcolor=darkred
            ]{hyperref}

\newcommand{\affA}{State Key Laboratory of Artificial Microstructure and Mesoscopic Physics, School of Physics, Peking University, 100871 Beijing, China}
\newcommand{\affB}{Department of Physics, 
Institute of Nanotechnology and Advanced Materials, 
Bar-Ilan University, Ramat-Gan 52900, Israel}

\begin{document}

\title{Preparation of cat states in many-body eigenbasis via non-local measurement}
\author{Ruoyu Yin}
\affiliation{\affA}
\affiliation{\affB}
\author{Hongzheng Zhao}
\email{hzhao@pku.edu.cn}
\affiliation{\affA}


\begin{abstract}
Engineered dissipation offers a promising route to prepare correlated quantum many-body states that are otherwise difficult to access using purely unitary protocols. However, creating superpositions of multiple many-body eigenstates with tunable properties remains a major challenge.
We propose to periodically interrupt the many-body evolution by precisely removing a given many-body Fock state through a non-local post-selected measurement protocol. Upon tuning the measurement period, we show that a dark state manifold survives the removal, allowing us to filter the system and generate a coherent superposition within this manifold at long times. 
As a testbed, we study a non-integrable spin-1 XY chain featuring a solvable family of eigenstates that can differ macroscopically in quasi-particle excitations. Our protocol generates tunable superpositions of these eigenstates, including the spin-1 Greenberger–Horne–Zeilinger state and a generalized variant with tunable spatiotemporal order. Under perturbations, the system exhibits an exceptionally long-lived metastable regime where the engineered superpositions remain robust. Our work provides new insight into quantum state preparation via non-local measurements using tools available in current quantum simulators.
\end{abstract}

\maketitle

{\em Introduction.---}
Quantum state preparation for many-body systems is of fundamental significance 
in quantum computing and simulation \cite{Seth1996,Cirac2012,Biamonte2017,Markus2017,Cao2019,Sam2020}. 
There are generally two broad categories of state preparation schemes:
Purely unitary protocols include time-dependent adiabatic algorithms~\cite{farhi2000adiabatic,Greiner2002,Albash2018adiabatic,claeys2019floquet,Lukin2021adiabatic}
and variational quantum eigensolvers~\cite{Cerezo2021,Tilly2022variational,Fedorov2022variational};  
In contrast, non-unitary protocols leverage engineered reservoir~\cite{Zoller1996,Giovanna2007,Diehl2008,Verstraete2009,Xiaoting2010,Cirac2011a,Barreiro2011,Shankar2013,Lin2013,Giovanna2015,Reiter2016,Xiaoting2016,Yuval2020Steer,Soonwon2021,Rall2023thermalstate,chen2023thermal,Takashi2023thermal} 
or coupling to dissipative auxiliary qubits~\cite{shtanko2021thermal,Giovanna2024quantumstate,Matthies2024programmable,Xiao2024Google,Xiao2025},
achieved through local measurements, post-selection, or dissipation channels~\cite{Zoller2008pra,Verstraete2009,Reiter2016,lu2023mixed,lavasani2023monitored,ritter2024autonomous,schnell2024dissipative,zhao2025feedback}.
Consequently, the system can be steered into
ground states~\cite{Zoller2008pra,Xiao2025},
finite-temperature thermal states
\cite{shtanko2021thermal,Rall2023thermalstate,chen2023thermal,lin2025dissipative},
or non-stationary states using dynamical symmetry \cite{Buca2019,Nishant2019,Buca2023prx}.

In this work, we investigate how to prepare {\it coherent} superpositions of multiple many-body eigenstates, e.g., a cat state in the energy basis, through engineered dissipation. We ask whether or not, and under which conditions,
one can tune the key features of this superposition, such as the number of constituent eigenstates and their relative phases, using the toolset provided by current quantum simulators. Note that this target is fundamentally different from preparing easily accessible product states, which normally overlap with exponentially many eigenstates, a fact that severely constrains their tunability in the energy basis. Realizations of such cat states have attracted long-standing interest both in fundamental physics \cite{GHZ1990,Horodecki2009RMP,Benami2025}, quantum computing and metrology \cite{Reed2012,Toth2014}. A paradigmatic example is the Greenberger–Horne–Zeilinger (GHZ) state, for which many preparation schemes have by now been proposed and experimentally implemented \cite{Pan1999,DiCarlo2010,Monz2011,Nishimori}. Yet, none provides an answer to our main question and applies to generic quantum many-body systems. 

Preparation of such superpositions is a demanding challenge for the following reasons:
(1) Eigenstates of generic quantum many-body systems are highly entangled, the exact determination of which can be costly in large systems.
(2) According to the eigenstate thermalization hypothesis (ETH)~\cite{d2016quantum}, eigenstates that are close in energy are locally indistinguishable. Thus, local dissipation or measurements acting uniformly within a narrow energy shell are insufficient for our purpose.

\begin{figure}[t]
\centering
\includegraphics[width=\linewidth]{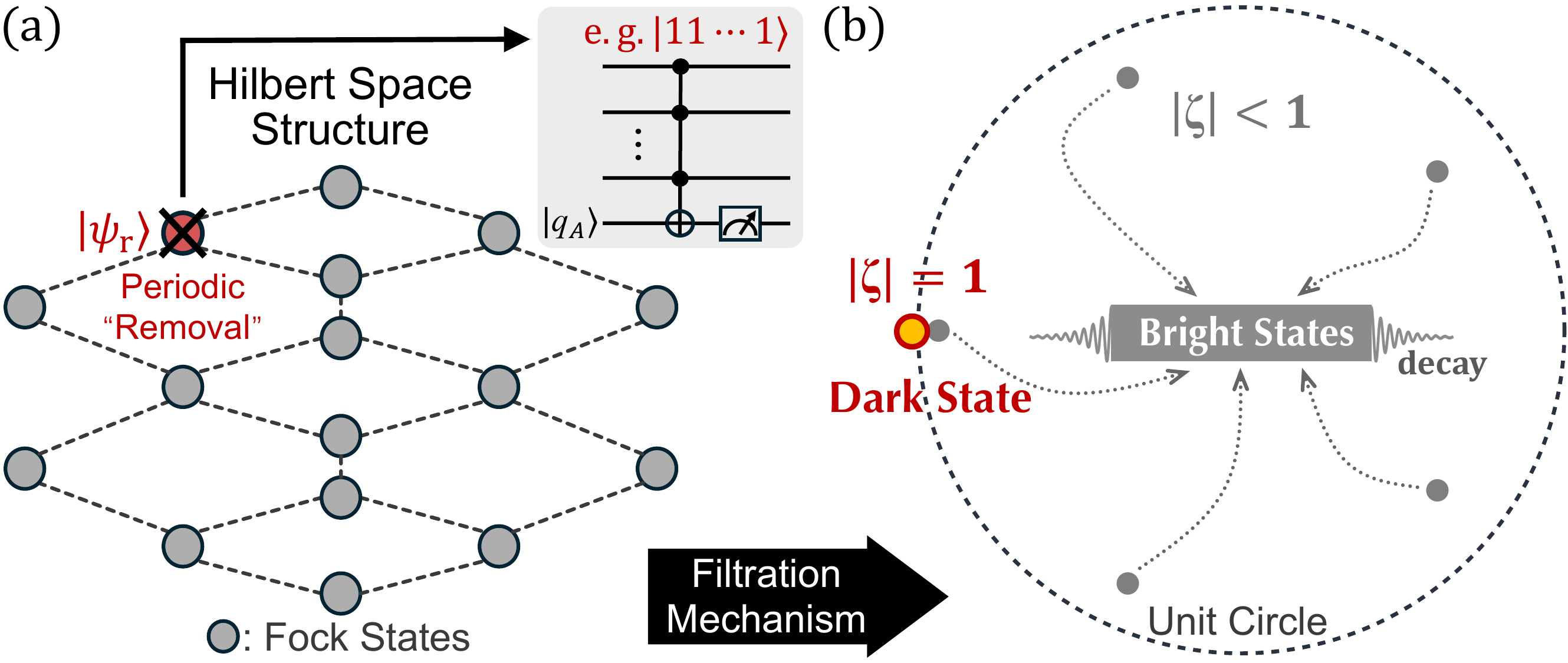}
\caption{(a) The state filtration protocol periodically removes a many-body Fock state $\ket{\psi_{\rm r}}$. This can be realized by coupling the system to one ancilla with a multi-body controlled gate. The measurement outcome of the ancilla is then used to post-select desired trajectories. 
(b) The undetectable dark states remain invariant under the filtration,
while the bright states are depleted at long times.
}
\label{fig:protocol}
\end{figure}

Here, we propose a way to overcome these challenges
and construct a generic protocol to prepare superpositions of many-body eigenstates. It uses only a single ancillary qubit and does not require prior
knowledge of the explicit form of the eigenstates, or any
symmetry structure of the underlying Hamiltonian. The key conceptual ingredient is
a novel post-selected measurement scheme, which removes one specific many-body Fock state while maintaining the coherence of the rest of the system (Fig.~\ref{fig:protocol}(a)). Therefore, this protocol is sharply distinct from the conventional projective measurement, which can collapse the entire wavefunction completely.  
We use this measurement protocol to periodically interrupt a unitary evolution, $\hat{U}(\tau) = e^{-iH\tau}$ with a many-body Hamiltonian $H$, at the measurement period $\tau$. 
{A suitable choice of $\tau$ leads to many-body resonances between the target eigenenergies, creating an undetectable dark state manifold that survives the periodic removal.
Consequently, for a simple product state as the initial state that has a finite probability in this manifold, this protocol effectively ``filtrates'' the system by repeatedly removing other detectable states, eventually generating a coherent superposition within this manifold (Fig.~\ref{fig:protocol}(b))}.

Similar protocols have primarily been studied in single-particle systems using local-in-space measurements, leading to important discoveries related to the quantum first-detection time and its statistical properties~\cite{Krovi2006,Gruenbaum2013,Dhar2015,Friedman2017a,Lahiri2019,dubey2021quantum,Sabine2022,Ruoyu2023,Yajing2023,Zhenbo2023,yin2024restart,Giovanna2025causality}. 
{Here, we propose to use many-body controlled gates and a single ancillary qubit to realize such non-local measurements in quantum many-body systems (Fig.~\ref{fig:protocol}(a)).}

While this method generally applies to generic many-body systems, we find the spin-1 XY model an ideal setting for demonstration as it provides a clear physical interpretation of the dark state manifold that we can engineer. Across the entire many-body spectrum, this model features a subset of exactly solvable eigenstates which contain a different number of quasi-particles~\cite{Thomas2019,Bernevig2018a,Sanjay2020,Mark2020a,Roderich2023Review}. We provide explicit examples and demonstrate how to engineer the many-body resonance,
and hence the dark states, by tuning the measurement period. Concretely, we first restrict the system's evolution within the solvable subspace and periodically remove a simple product state. We demonstrate the controllability of the resulting superpositions by first generating a spin-1 GHZ state;
going beyond, we also generate a dynamical cat state exhibiting spatiotemporal order, whose concrete form relies on both Hamiltonian parameters and the initial state. Remarkably, both states are composed of multiple many-body eigenstates which can differ macroscopically in the number of quasi-particles, and hence are generally unreachable in a thermal equilibrium. 
Crucially, in the presence of perturbations, we identify an exceptionally long-lived metastable regime where our protocol remains robust, thereby rendering the realization of our protocol feasible in practice. Possible experimental realization
will be discussed before concluding.

{\em Measurement-induced state filtration protocol.---}
We consider a system governed by the Hamiltonian $H$ 
and initialized in 
$\ket{\psi_0}$. 
In a stroboscopic manner with period $\tau$, 
we interrupt the unitary evolution of the system with measurements,
attempting {\em not} to detect it in a predetermined
state $\ket{\psi_{\rm r}}$. 
This non-detection measurement is presented by $\mathbbm{1}-|\psi_{\rm r}\rangle \langle\psi_{\rm r}|$ 
($\mathbbm{1}$ is the identity matrix),
removing the state $\ket{\psi_{\rm r}}$ from the wavefunction
(see Fig. \ref{fig:protocol}(a))
\cite{walter2025thermodynamic}.
Practically, this can be realized using post-selections \cite{GoogleQuantumAI2023,koh2023measurement}.
Following $n$ consecutive conditional measurements with non-detection outcomes,
the state of the system becomes
\begin{equation}\label{eq:generalformula}
\begin{aligned}
    \ket{\psi_n} &= {\rm N}_n (\mathcal{F}^\tau_{\psi_{\rm r}})^n \ket{\psi_0}, \\
    \text{with} \,\,\,
    {\cal F} ^\tau_{\psi_{\rm r}} 
    &= 
    \left(
    \mathbbm{1} - |\psi_{\rm r}\rangle \langle \psi_{\rm r}| 
    \right) U(\tau),
\end{aligned}
\end{equation}
where ${\rm N}_n$ is the normalization factor that depends on time $n$, and ${\cal F} ^\tau_{\psi_{\rm r}}$ generates the time evolution and filtrates the state.
Intuitively, the wavefunction continuously leaks out of the system due to the periodic removal.
Yet, one can engineer dark states that survive such periodic removal,
hence their name, through properly choosing a suitable measurement period $\tau$ \cite{Thiel2020D,Liu2022a}.
Formally, these dark states, $\ket{\Phi_{\delta}}$, satisfy the eigenfunction ${\cal F}^\tau_{\psi_{\rm r}} \ket{\Phi_{\delta}} = e^{-iE_{\delta}\tau} \ket{\Phi_{\delta}},$
with the corresponding eigenvalues $e^{-iE_{\delta}\tau}$ of unit modulus. All other eigenstates are dubbed bright states, whose eigenvalues have modulus smaller than 1,
$0\le|\zeta|<1$,
hence decaying under the {filtration} protocol
(Fig. \ref{fig:protocol}(b)).

The presence of dark states necessitates degeneracy in $U(\tau)$.
For example, consider two eigenstates $|E_2\rangle$ and $|E_1\rangle$ of the Hamiltonian $H$ with $E_1\neq E_2$. We choose $\tau$ s.t. the following resonance condition is satisfied~\cite{yin2019}
\begin{equation}
\label{eq:degeneracy}
    e^{-i E_1 \tau}= e^{-i E_2 \tau}.
\end{equation}
This leads to the dark state 
$|\Phi\rangle = {\rm N} 
( 
\langle \psi_{\rm r}|E_2\rangle |E_1\rangle - 
\langle \psi_{\rm r}|E_1\rangle |E_2\rangle 
)$,
satisfying
${\cal F}^\tau_{\psi_{\rm r}} \ket{\Phi} {=} e^{-iE_1\tau} \ket{\Phi}$ with the eigenvalue $e^{-i E_1 \tau}$, {without {\it a prior} knowledge of the concrete form of the eigenstates.}
More generally, for cases where an eigenvalue of \(U(\tau)\) has degeneracy $g>2$, 
$g{-}1$ dark states can be constructed using the Gram-Schmidt procedure~\cite{Thiel2020D},
see details in the End Matter.
Importantly, the resonance condition implies that the relevant eigenstates acquire the same phase under $U(n\tau)$. Thus $\ket{\Phi}$ is stationary under $U(n\tau)$, i.e., $U(n\tau) \ket{\Phi} {=} e^{-i n\tau E_1} \ket{\Phi}$, guaranteeing
that $\ket{\Phi}$ remains orthogonal to $\ket{\psi_{\rm r}}$ at stroboscopic times $t{=}n\tau$.
Thus, through engineering the resonance or degeneracy of $U(\tau)$
we construct non-detectable subspaces spanned by the degenerate eigenstates.
Note, the presence of symmetry and integrability may further induce degeneracies in the spectrum of $H$, which complicate our analysis. 
Here, for simplicity, we always assume the spectrum of $H$ does not have any degeneracy.

By decomposing the wavefunction into dark and bright states, Eq.~\eqref{eq:generalformula} can be rewritten as
\begin{equation}\label{eq:psi_n_decomposition}
    \ket{\psi_n} =  {\rm N}_n \Big(\sum_{|\zeta_k|<1} \eta_{\zeta_k} \ket{\zeta_k^r} \zeta_k^n 
                    + 
                    \sum_{\delta} \eta_{\delta} \ket{\Phi_{\delta}} e^{-in E_{\delta} \tau}\Big),
\end{equation}
where $\eta_{\zeta_k}$ quantifies the overlap between the initial states and bright/dark states,
$\eta_{\zeta_k} = \langle \zeta_k^l | \psi_0 \rangle/\langle \zeta_k^l | \zeta_k^r \rangle$,
with $\bra{\zeta_k^l}$ ($\ket{\zeta_k^r}$) denoting the left (right) eigenvector of
${\cal F}^\tau_{\psi_{\rm r}}$ with the eigenvalue $\zeta_k$.
For dark states, the left and right eigenvectors are Hermitian conjugates.
Hence, in the long-time limit we obtain
\begin{equation}\label{destiny}
\begin{aligned}
\lim_{n\to\infty} \ket{\psi_n} =
\mathrm{N}_{\infty} \sum_{\delta} e^{-i n \tau E_{\delta}} 
\langle \Phi_\delta | \psi_0 \rangle | \Phi_\delta \rangle.
\end{aligned}
\end{equation}
A steady state arises if only one dark state $\ket{\Phi_\delta}$ satisfying $\eta_{\delta}=\langle \Phi_\delta | {\psi_0} \rangle \neq0$, 
while other overlaps $\eta_{\delta}=0$.
If multiple dark states $\ket{\Phi_\delta}$ have nonzero overlaps, a superposition of multiple dark states can be generated.

This scheme generally applies to many-body interacting systems, as we numerically verified in the End Matter using a toy model, a random matrix Hamiltonian, which is deemed to capture the most generic features in many-body interacting systems~\cite{d2016quantum}.
However, Eq.~\eqref{eq:degeneracy} requires the knowledge of the exact eigenvalues, which can be difficult to obtain in generic non-integrable many-body systems. Also, eigenstates of a random matrix are mostly featureless, as is the resulting superposition after the filtration procedure. Therefore, in the following we focus on the spin-1 XY model, which, despite being non-integrable, features a family of exactly solvable eigenstates. These eigenstates have a clear physical interpretation in terms of quasi-particle excitations. Eventually, our protocol leads to a superposition of eigenstates that can differ macroscopically in the number of quasi-particles, which is difficult to obtain in thermal equilibrium.

{\em The model.---}
We consider the Hamiltonian
\begin{equation}\label{ham}
    H = 
    J \sum_{i}^L 
    \left(
    S_i^x S_{i+1}^x 
    + S_i^y S_{i+1}^y 
    \right) 
    + h \sum_i S_i^z 
    + D\sum_i (S_i^z)^2,
\end{equation}
where $i$ labels the lattice sites
and the operator $S_i^\alpha (\alpha = x,y,z)$ are spin-$1$ operators.
The spin-1 degrees of freedom are denoted by $\ket{\pm_i}$, $\ket{0_i}$, 
corresponding to the eigenvalues $\pm 1$ and $0$ under $S_i^z$, respectively.
Open boundary conditions are used here. 
This model is non-integrable and most of the eigenstates obey ETH. However, there exists a family of solvable eigenstates that we use to engineer dark states. To see this, we start from the vacuum
$\ket{\Omega} =\bigotimes_i^L |-_i\rangle$, 
and apply the operator $Q^+ = (1/2)\sum_{j} e^{i\pi j} (S_j^+)^2$ to inject one 
``bi-magnon'' quasi-particle excitation of momentum $\pi$,
as the bound state of two magnon excitations.
Interestingly, as shown in Ref.~\cite{Sanjay2020,Mark2020b},
starting from $\ket{\Omega}$, the following spectrum-generating algebra (SGA) can be obtained
\begin{equation}
\label{eq:SpinRSGA}
    [H, Q^+] \ket{\Omega} = 2h Q^+ \ket{\Omega},
\end{equation} leading to a tower of solvable non-ergodic eigenstates 
\begin{equation}
\label{eq:bimagnon}
    |{\cal B}_n\rangle = {\rm N}(Q^+)^n \ket{\Omega},
\end{equation} 
where \({\rm N}\) ensures normalization. $|{\cal B}_n\rangle$ contains $n$ bi-magnons with eigenenergy \(E_n = E_0 + 2nh\), where $E_0 = (D-h)L$. 
The SGA structure is preserved upon adding $H_3= J_3 \sum_{i}^L (S_i^x S_{i+3}^x + S_i^y S_{i+3}^y )$ to $H$.
Hereinafter, we fix $J=1,D=0.1$.

\begin{figure}[tp]
\centering
\includegraphics[width=\linewidth]{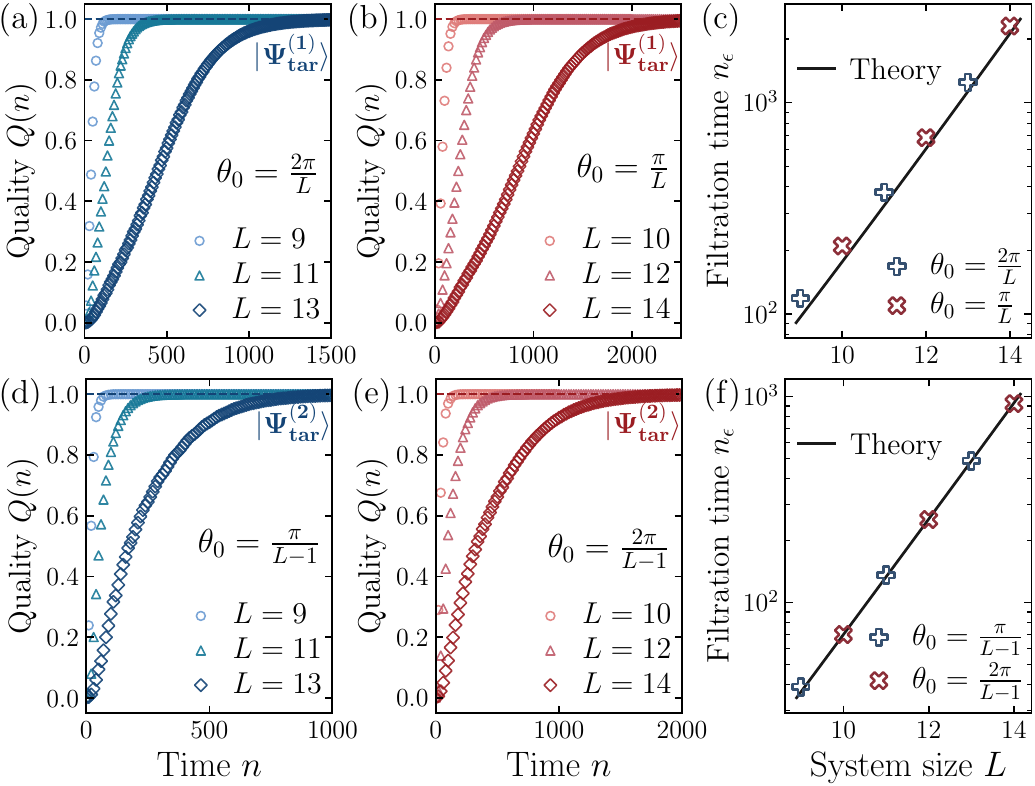}
\caption{
The filtration quality $Q_n$ for $\ket*{\Psi_{\rm tar}^{(1)}}$ (a,b)
or $\ket*{\Psi_{\rm tar}^{(2)}}$ (d,e) 
as a function of the number of measurements $n$. 
The model here is the spin-1 XY chain (\ref{ham}) with $h=1$, $J_3=0.1$.
The initial state is parameterized by $\theta_0$, set as different values shown in the figure.
The removal state is the product state (\ref{neel}) as mentioned.
(c,f) The {filtration time}, $n_\epsilon$, required to reach the fidelity $Q_n = 1-\epsilon$,
exhibits an exponential law of the system size $L$. Here $\epsilon=0.01$.
The numerical results (crosses) fit nicely with the theory (black lines).
}
\label{fig:QnGHZ}
\end{figure}

We choose a simple product state
as the removal state $\ket{\psi_{\rm r}}$ for the filtration process,
\begin{equation}\label{neel}
    \ket{\psi_{\rm r}}
    =\bigotimes_j 
    \left(
    {\left|+_j\right\rangle
    -
    e^{i\pi j}\left|-_{j}\right\rangle}
    \right)/{\sqrt{2}}.
\end{equation}
The initial state is chosen as
\begin{equation}\label{ini}
    \ket{\psi_0} 
    = \bigotimes_j 
    \left(
    {\left|+_j\right\rangle
    +
    e^{i(j\pi + \theta_0)} \left|-_{j}\right\rangle}
    \right)
    /{\sqrt{2}},
\end{equation}
where we explicitly introduce the parameter $\theta_0$ to manipulate the relative phase in the superposition.
Also, both the initial and removal states live within the solvable subspace, as one can decompose them as $\ket{\psi_{\rm r}} =\sum_{n=0}^L e^{in\pi} \sqrt{\tbinom{L}{n}  / 2^{L}} \left|{\cal B}_n\right\rangle$, and 
$\ket{\psi_0}  = 
\sum_{n=0}^L e^{i(L-n)\theta_0}
\sqrt{\tbinom{L}{n} / 2^{L}} \left|{\cal B}_n\right\rangle$. Therefore, the system dynamics under filtration is restricted inside this non-ergodic subspace, making our theoretical analysis of the dark state manifold particularly simple.

Without loss of generality, we consider two concrete values of $\tau$ to engineer the dark states and the eventual target states.
First, when $h\tau_1 = \pi/L$,
the only resonance is $e^{-i\tau_1 E_0} = e^{-i\tau_1 E_L}$ and the only dark state reads
\begin{equation}\label{tar1}
\begin{aligned}
    \ket*{\Psi_{\rm tar}^{(1)}} &= 
    \left( \ket{{\cal B}_L} - e^{iL\pi}\ket{{\cal B}_0} \right)/ \sqrt{2}
    \\
    &=
    \left(|++\cdots\rangle - e^{iL\pi}|--\cdots\rangle \right)/ \sqrt{2}.
\end{aligned}
\end{equation}
It corresponds to a static superposition
of two macroscopically distinct ferromagnetic configurations, or the spin-1 GHZ state. It exhibits a long-range spatial order since the expectation value of the string operator
$\prod_i X_i$, where $X_i$ directly flips each spin between $\ket{+_i}$ and $\ket{-_i}$, is a nonzero constant regardless of the system size.

This can be generalized such that multiple dark states survive the removal. For instance, when choosing $h\tau_2 = \pi/(L-1)$ we obtain two dark states, $\ket{\Phi_1} 
= {- 1 \over \sqrt{L+1}} \left[ (-1)^L \sqrt{L} \ket{{\cal B}_0} + \ket{{\cal B}_{L-1}}\right]$,
and $\ket{\Phi_2} = 
{1 \over \sqrt{L+1}} \left[ (-1)^L \ket{{\cal B}_1} + \sqrt{L} \ket{{\cal B}_{L}}\right]$. At long times the system indeed forms a superposition between these two
\begin{equation}\label{tar2}
\begin{aligned}
    \ket*{\Psi_{\rm tar}^{(2)}} =    
                                \left(
                                \ket{\Phi_1} e^{i\theta_0} 
                                -  
                                \ket{\Phi_2} 
                                e^{-i n \tau_2 \Delta E_{01} }
                                \right)/\sqrt{2}, 
\end{aligned}
\end{equation}
which evolves in time with an oscillating frequency $\Delta E_{01}= E_1-E_0 = 2h$.
{It is worth highlighting that, both examples induce a coherent superposition between eigenstates that can differ macroscopically in the number of bi-magnon excitations. More interestingly, $\ket*{\Psi_{\rm tar}^{(2)}}$ further demonstrates spatiotemporal order, which is absent in the static case, $\ket*{\Psi_{\rm tar}^{(1)}}$. This becomes manifest in the expectation value of the string operator,
$\expval{\prod_i X_i} = \cos(\theta_0 + 2n h\tau)$, {offering a direct experimental dynamical signature of the successful preparation of the cat state~\cite{Lukin2021adiabatic}.}
Phase of this oscillation also explicitly depends on the angle $\theta_0$, precisely tunable by the initial states, see Eq.~\eqref{ini}.

To quantify how closely the system approaches the target state,
we use the state fidelity
\begin{equation}\label{Qn}
    Q_n:=
    | \langle \Psi_{\rm tar} | \psi_n \rangle |^2,
\end{equation}
where the target state $\ket{\Psi_{\rm tar}}$ 
can be either $\ket*{\Psi_{\rm tar}^{(1)}}$ or $\ket*{\Psi_{\rm tar}^{(2)}}$.
As verified in Fig.~\ref{fig:QnGHZ}}, starting from 0, $Q_n$ approaches unity in the long time limit, confirming the validity of our protocol. Also, for larger system sizes, it generally takes longer to reach high fidelity. To quantify this, we fix a small tolerance value $\epsilon$ and extract {the filtration time} $n_\epsilon$ as the smallest integer $n$ such that $Q_{n} \ge 1-\epsilon$.
As shown in Fig.~\ref{fig:QnGHZ} (c) and (f), for both target states $n_\epsilon$ scales exponentially in $L$.

\begin{figure}[t]
\centering
\includegraphics[width=\linewidth]{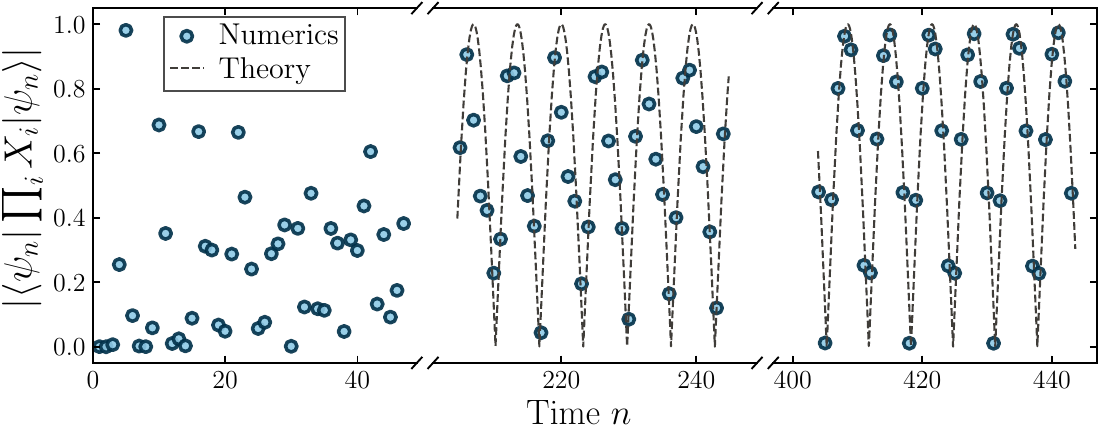}
\caption{The oscillatory behavior of $|\langle \psi_n | \prod_i X_i | \psi_n \rangle|$
as a function of measurement time $n$.
The model and parameters are the same as in Fig. \ref{fig:QnGHZ}(e),
with $L=14$.
For small $n$, irregular dynamics in the expectation value of the string operator is observed.
Periodic oscillation gradually appears and converges to the theoretical prediction as $n$ increases.
We note that observing such oscillatory behavior requires significantly fewer measurements than those needed to achieve high $Q_n$.
}
\label{fig:ParityL14}
\end{figure}

The convergence of $Q_n$ is governed
by the dominant bright-state eigenvalue, denoted by $\zeta_d$,
whose modulus is closest to $1$ and hence decays the slowest. By mapping the bright-state eigenvalues to an electrostatic system (see Note 1 in the Supplementary Material (SM)) and using perturbative analysis \cite{Gruenbaum2013,yin2019}, we obtain the analytical expressions of $n_{\epsilon}$ (black lines in Fig.~\ref{fig:QnGHZ} (c) and (f)), which match well with the numerics,
and are detailed in SM Note 2.
We find that
$|\zeta_d|^2 \propto \exp\left(-\sum_k |\langle {\cal B}_k | \psi{\rm _r} \rangle|^2\right)$, where ${\ket{{\cal B}_k}}$ span the dark subspace.
Since $|\langle {\cal B}_k | \psi_{\rm r} \rangle|^2 \propto 2^{-L}$,
the depletion time of the dominant bright state grows exponentially with $L$.

We note that for other target states constructed by tuning $\tau$ to different resonance conditions, $n_{\epsilon}$ can have different scaling law versus $L$, e.g., it may decrease for larger $L$, see details in the End Matter.

Through measuring the string operator one can also verify the successful preparation of those states. As shown in Fig.~\ref{fig:ParityL14}, at early times the expectation value, $\expval{\prod_i X_i}$, exhibits irregular dynamics. Periodic oscillation starts to appear around $n \approx 200$, where the system enters the dark state manifold spanned by 
$\{\ket{{\cal B}_0},\ket{{\cal B}1},\ket{{\cal B}_{L-1}},\ket{{\cal B}_L}\}$. Then the protocol further steers the system to the eventual target state, and $\expval{\prod_i X_i}$ accurately follows the theoretical prediction (dashed line) for $n \geq 400$. 
Therefore, observation of the oscillatory string operator requires far fewer measurements than those needed to achieve high target fidelity, cf. Fig.~\ref{fig:QnGHZ}(e) where $Q_n=0.99$ when $n\approx 2000$.

\begin{figure}[t]
\centering
\includegraphics[width=\linewidth]{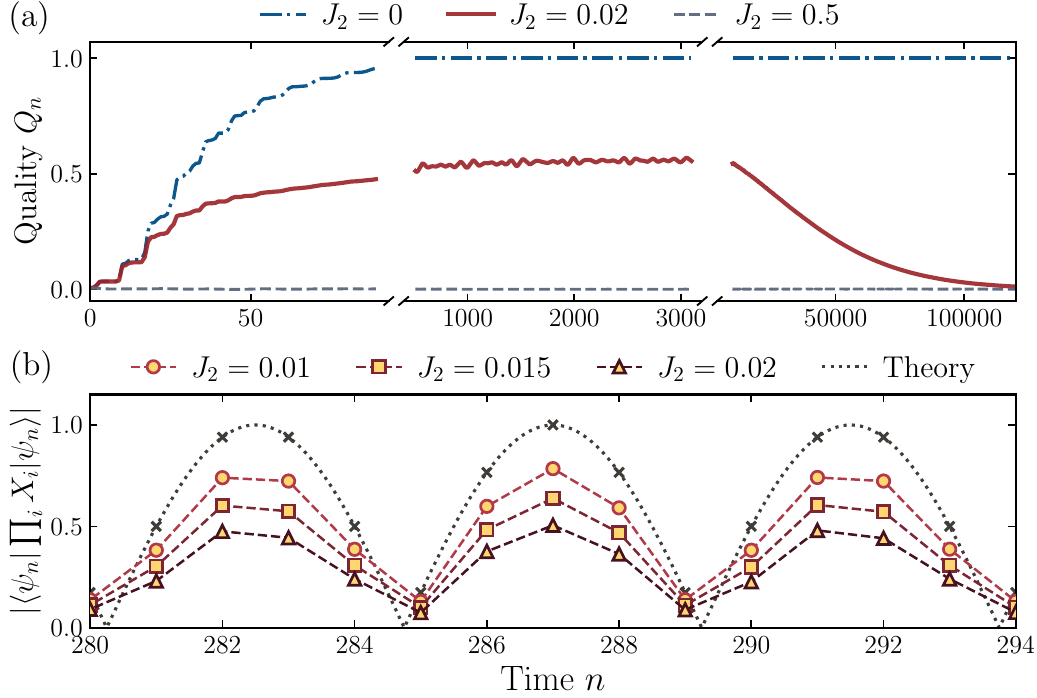}
\caption{(a) The filtration quality $Q_n$ in the presence of perturbations.
For numerical efficiency, we set $\lambda=0$, $J_3=0$.
(b) The oscillatory dynamics of the string operator persists in the presence of perturbations.
Here we set $\lambda=0.01$.
Other parameters are as in Fig. \ref{fig:QnGHZ}(e) with $L=10$.
Weak perturbations lead to a long-lived plateau of $Q_n$ and allow the observation of oscillatory $\expval{\prod_i X_i}$ in a long time window.
}
\label{fig:flipL10}
\end{figure}

{\em Stability of the protocol.---}
We now demonstrate the stability of our protocol against generic perturbations.
First, we add a vector $\lambda \ket{\nu}$ to the removal state $|\psi_{\rm r} \rangle$, where $\ket{\nu}$ is a random vector in the total Hilbert space and $\lambda$ quantifies its strength. This mimics the imperfection during the removal process.
Then we introduce a next-nearest-neighbor XY exchange term to Eq.~\eqref{ham},
$H_2 = J_2\sum_{i}^L \left(S_i^x S_{i+2}^x + S_i^y S_{i+2}^y \right)$, which breaks SGA structure. Therefore, most eigenstates constructed by Eq.~\eqref{eq:bimagnon} are no longer solvable as they couple to other thermal eigenstates. One exception arises where $\ket{{\cal B}_0}$ and $\ket{{\cal B}_L}$ still remain eigenstates of the system, and hence $\ket*{\Psi_{\rm tar}^{(1)}}$ (Eq.~\eqref{tar1}) can still be a dark state and remain stable at long times. In contrast, the second target state, $\ket*{\Psi_{\rm tar}^{(2)}}$, is unstable and the state-fidelity $Q_n$ drops to zero in the long time limit, as verified in Fig.~\ref{fig:flipL10}(a). 

It is worth noting that, a transient but exceptionally long-lived metastable regime appears for weak perturbations. As shown in Fig.~\ref{fig:flipL10}(a), for $J_2=0.02$ (red line), $Q_n$ initially follows the unperturbed filtration (blue dashed line) and gradually enters a plateau with $Q_n\approx 0.5$. This metastable plateau persists until $n\approx 10^4$ before the onset of the slow damping process. 
Therefore, despite the coupling to other thermal eigenstates, the original dark state manifold still dominates the early-time evolution and steers the state properly as desired. This provides a sufficiently long time window where the coherent oscillatory dynamics of the string operator is clearly visible, as shown in Fig.~\ref{fig:flipL10}(b), despite that the oscillation amplitude decreases linearly for larger values of perturbation. 

{\em Possible experimental implementation.---}
This spin-1 XY model is readily implementable in existing quantum simulation experiments~\cite{senko2015realization}. Exact or approximate SGA structure has also been identified in a variety of quantum many-body systems~\cite{Sanjay2020,Mark2020a,Odea2020,Roderich2023Review}, e.g., the PXP model which effectively captures Rydberg systems in the strong Rydberg blockade regime \cite{Lukin2017Rydberg}.
Crucially, the state filtration protocol is physically implementable with the aid of a single ancillary qubit. One can couple the many-body system to this ancilla using a multi-body controlled gate~\cite{Shor1995Elem,shi2002both,Shende2009Toffoli,wu2025engineering}, such that only when the system is in the product state $\ket{\psi_{\rm r}}$ (Eq.~\eqref{neel}), the ancilla can flip. Through directly measuring the ancilla, one can repeatedly post-select trajectories
where the system avoids detection in state $\ket{\psi_{\rm r}}$. {Realization of such multi-body controlled gates, e.g. the $n-$body Toffoli gate, can be efficient~\cite{Monz2009RealizationToffoli,Evered2023HighFid,cao2024multi}, using $\mathcal{O}(\log n)$-depth quantum circuits~\cite{nie2024quantum} or Floquet engineering without gate decomposition~\cite{wu2025engineering}.}

{\em Discussions.---}
We propose to generate a superposition of multiple many-body eigenstates within a dark state manifold, by periodically removing a given many-body Fock state. The key ingredient in realizing the dark state is to engineer degeneracies in $U(\tau)$ by tuning $\tau$. This strategy is independent of the microscopic details of the underlying Hamiltonian, and does not require any symmetry structure as in other state-preparation protocols~\cite{Buca2019,Huangbiao2024nc}. Existing experimental toolset suffices to realize the filtration process, thus paving the way to engineer dissipation and prepare novel quantum states with non-local measurements~\cite{Thomas2025PRXQ}.

We demonstrate this protocol by focusing on the spin-1 XY model with the SGA structure and generating cat states over exactly solvable eigenstates with macroscopically different numbers of quasi-particles. 
More examples can be found in the End Matter.
We emphasize that these discussions are not limited to this specific system, and one can readily generate superpositions of different types of non-ergodic many-body eigenstates, such as the ones with volume-law entanglement~\cite{Zhao2021Orthogonal,Langlett2022volume,mukherjee2025symmetrictensorscarstunable}. 

Perturbations that explicitly break the SGA lead to a transient yet exceptionally long-lived regime where our protocol remains stable. Phenomenologically, this long-lived behavior is reminiscent of the prethermalization phenomena observed in Floquet systems~\cite{ho2023quantum}, and it
is intriguing to further build up their connection in future studies. 
Moreover, generalizing the current periodic protocol to non-periodic~\cite{zhao2021random,pilatowsky2025critically} or adaptive schemes~\cite{lu2023mixed,pocklington2025accelerating} to improve the depletion efficiency is worth pursuing.

This work is closely related to the quantum walk recurrence time,
topologically tied to the dimension of Hilbert space reachable from the initial state \cite{Gruenbaum2013,Sinkovicz2015}.
Our methods and results thus offer a framework for exploring these intriguing questions in quantum many-body systems.

{\em Acknowledgments.---}
We thank Roderich Moessner, Tianfeng Feng, Markus Heyl, Eli Barkai, and Jie Ren for enlightening discussions.
This work is supported by the National Natural Science Foundation of China (Grant No. 12474214),
and by Innovation Program for Quantum Science and Technology (No. 2024ZD0301800),
and by ``The Fundamental Research Funds for the Central Universities, Peking University'',
and by ``High-performance Computing Platform of Peking University".


\begin{thebibliography}{101}%
\makeatletter
\providecommand \@ifxundefined [1]{%
 \@ifx{#1\undefined}
}%
\providecommand \@ifnum [1]{%
 \ifnum #1\expandafter \@firstoftwo
 \else \expandafter \@secondoftwo
 \fi
}%
\providecommand \@ifx [1]{%
 \ifx #1\expandafter \@firstoftwo
 \else \expandafter \@secondoftwo
 \fi
}%
\providecommand \natexlab [1]{#1}%
\providecommand \enquote  [1]{``#1''}%
\providecommand \bibnamefont  [1]{#1}%
\providecommand \bibfnamefont [1]{#1}%
\providecommand \citenamefont [1]{#1}%
\providecommand \href@noop [0]{\@secondoftwo}%
\providecommand \href [0]{\begingroup \@sanitize@url \@href}%
\providecommand \@href[1]{\@@startlink{#1}\@@href}%
\providecommand \@@href[1]{\endgroup#1\@@endlink}%
\providecommand \@sanitize@url [0]{\catcode `\\12\catcode `\$12\catcode
  `\&12\catcode `\#12\catcode `\^12\catcode `\_12\catcode `\%12\relax}%
\providecommand \@@startlink[1]{}%
\providecommand \@@endlink[0]{}%
\providecommand \url  [0]{\begingroup\@sanitize@url \@url }%
\providecommand \@url [1]{\endgroup\@href {#1}{\urlprefix }}%
\providecommand \urlprefix  [0]{URL }%
\providecommand \Eprint [0]{\href }%
\providecommand \doibase [0]{https://doi.org/}%
\providecommand \selectlanguage [0]{\@gobble}%
\providecommand \bibinfo  [0]{\@secondoftwo}%
\providecommand \bibfield  [0]{\@secondoftwo}%
\providecommand \translation [1]{[#1]}%
\providecommand \BibitemOpen [0]{}%
\providecommand \bibitemStop [0]{}%
\providecommand \bibitemNoStop [0]{.\EOS\space}%
\providecommand \EOS [0]{\spacefactor3000\relax}%
\providecommand \BibitemShut  [1]{\csname bibitem#1\endcsname}%
\let\auto@bib@innerbib\@empty
\bibitem [{\citenamefont {Lloyd}(1996)}]{Seth1996}%
  \BibitemOpen
  \bibfield  {author} {\bibinfo {author} {\bibfnamefont {S.}~\bibnamefont
  {Lloyd}},\ }\bibfield  {title} {\bibinfo {title} {Universal quantum
  simulators},\ }\href {https://doi.org/10.1126/science.273.5278.1073}
  {\bibfield  {journal} {\bibinfo  {journal} {Science}\ }\textbf {\bibinfo
  {volume} {273}},\ \bibinfo {pages} {1073} (\bibinfo {year}
  {1996})}\BibitemShut {NoStop}%
\bibitem [{\citenamefont {Cirac}\ and\ \citenamefont
  {Zoller}(2012)}]{Cirac2012}%
  \BibitemOpen
  \bibfield  {author} {\bibinfo {author} {\bibfnamefont {J.~I.}\ \bibnamefont
  {Cirac}}\ and\ \bibinfo {author} {\bibfnamefont {P.}~\bibnamefont {Zoller}},\
  }\bibfield  {title} {\bibinfo {title} {Goals and opportunities in quantum
  simulation},\ }\href {https://doi.org/10.1038/nphys2275} {\bibfield
  {journal} {\bibinfo  {journal} {Nature Physics}\ }\textbf {\bibinfo {volume}
  {8}},\ \bibinfo {pages} {264} (\bibinfo {year} {2012})}\BibitemShut {NoStop}%
\bibitem [{\citenamefont {Biamonte}\ \emph {et~al.}(2017)\citenamefont
  {Biamonte}, \citenamefont {Wittek}, \citenamefont {Pancotti}, \citenamefont
  {Rebentrost}, \citenamefont {Wiebe},\ and\ \citenamefont
  {Lloyd}}]{Biamonte2017}%
  \BibitemOpen
  \bibfield  {author} {\bibinfo {author} {\bibfnamefont {J.}~\bibnamefont
  {Biamonte}}, \bibinfo {author} {\bibfnamefont {P.}~\bibnamefont {Wittek}},
  \bibinfo {author} {\bibfnamefont {N.}~\bibnamefont {Pancotti}}, \bibinfo
  {author} {\bibfnamefont {P.}~\bibnamefont {Rebentrost}}, \bibinfo {author}
  {\bibfnamefont {N.}~\bibnamefont {Wiebe}},\ and\ \bibinfo {author}
  {\bibfnamefont {S.}~\bibnamefont {Lloyd}},\ }\bibfield  {title} {\bibinfo
  {title} {Quantum machine learning},\ }\href
  {https://doi.org/10.1038/nature23474} {\bibfield  {journal} {\bibinfo
  {journal} {Nature}\ }\textbf {\bibinfo {volume} {549}},\ \bibinfo {pages}
  {195} (\bibinfo {year} {2017})}\BibitemShut {NoStop}%
\bibitem [{\citenamefont {Reiher}\ \emph {et~al.}(2017)\citenamefont {Reiher},
  \citenamefont {Wiebe}, \citenamefont {Svore}, \citenamefont {Wecker},\ and\
  \citenamefont {Troyer}}]{Markus2017}%
  \BibitemOpen
  \bibfield  {author} {\bibinfo {author} {\bibfnamefont {M.}~\bibnamefont
  {Reiher}}, \bibinfo {author} {\bibfnamefont {N.}~\bibnamefont {Wiebe}},
  \bibinfo {author} {\bibfnamefont {K.~M.}\ \bibnamefont {Svore}}, \bibinfo
  {author} {\bibfnamefont {D.}~\bibnamefont {Wecker}},\ and\ \bibinfo {author}
  {\bibfnamefont {M.}~\bibnamefont {Troyer}},\ }\bibfield  {title} {\bibinfo
  {title} {Elucidating reaction mechanisms on quantum computers},\ }\href
  {https://doi.org/10.1073/pnas.1619152114} {\bibfield  {journal} {\bibinfo
  {journal} {Proceedings of the National Academy of Sciences}\ }\textbf
  {\bibinfo {volume} {114}},\ \bibinfo {pages} {7555} (\bibinfo {year}
  {2017})}\BibitemShut {NoStop}%
\bibitem [{\citenamefont {Cao}\ \emph {et~al.}(2019)\citenamefont {Cao},
  \citenamefont {Romero}, \citenamefont {Olson}, \citenamefont {Degroote},
  \citenamefont {Johnson}, \citenamefont {Kieferová}, \citenamefont
  {Kivlichan}, \citenamefont {Menke}, \citenamefont {Peropadre}, \citenamefont
  {Sawaya}, \citenamefont {Sim}, \citenamefont {Veis},\ and\ \citenamefont
  {Aspuru-Guzik}}]{Cao2019}%
  \BibitemOpen
  \bibfield  {author} {\bibinfo {author} {\bibfnamefont {Y.}~\bibnamefont
  {Cao}}, \bibinfo {author} {\bibfnamefont {J.}~\bibnamefont {Romero}},
  \bibinfo {author} {\bibfnamefont {J.~P.}\ \bibnamefont {Olson}}, \bibinfo
  {author} {\bibfnamefont {M.}~\bibnamefont {Degroote}}, \bibinfo {author}
  {\bibfnamefont {P.~D.}\ \bibnamefont {Johnson}}, \bibinfo {author}
  {\bibfnamefont {M.}~\bibnamefont {Kieferová}}, \bibinfo {author}
  {\bibfnamefont {I.~D.}\ \bibnamefont {Kivlichan}}, \bibinfo {author}
  {\bibfnamefont {T.}~\bibnamefont {Menke}}, \bibinfo {author} {\bibfnamefont
  {B.}~\bibnamefont {Peropadre}}, \bibinfo {author} {\bibfnamefont {N.~P.~D.}\
  \bibnamefont {Sawaya}}, \bibinfo {author} {\bibfnamefont {S.}~\bibnamefont
  {Sim}}, \bibinfo {author} {\bibfnamefont {L.}~\bibnamefont {Veis}},\ and\
  \bibinfo {author} {\bibfnamefont {A.}~\bibnamefont {Aspuru-Guzik}},\
  }\bibfield  {title} {\bibinfo {title} {Quantum chemistry in the age of
  quantum computing},\ }\href {https://doi.org/10.1021/acs.chemrev.8b00803}
  {\bibfield  {journal} {\bibinfo  {journal} {Chemical Reviews}\ }\textbf
  {\bibinfo {volume} {119}},\ \bibinfo {pages} {10856} (\bibinfo {year}
  {2019})}\BibitemShut {NoStop}%
\bibitem [{\citenamefont {McArdle}\ \emph {et~al.}(2020)\citenamefont
  {McArdle}, \citenamefont {Endo}, \citenamefont {Aspuru-Guzik}, \citenamefont
  {Benjamin},\ and\ \citenamefont {Yuan}}]{Sam2020}%
  \BibitemOpen
  \bibfield  {author} {\bibinfo {author} {\bibfnamefont {S.}~\bibnamefont
  {McArdle}}, \bibinfo {author} {\bibfnamefont {S.}~\bibnamefont {Endo}},
  \bibinfo {author} {\bibfnamefont {A.}~\bibnamefont {Aspuru-Guzik}}, \bibinfo
  {author} {\bibfnamefont {S.~C.}\ \bibnamefont {Benjamin}},\ and\ \bibinfo
  {author} {\bibfnamefont {X.}~\bibnamefont {Yuan}},\ }\bibfield  {title}
  {\bibinfo {title} {Quantum computational chemistry},\ }\href
  {https://doi.org/10.1103/RevModPhys.92.015003} {\bibfield  {journal}
  {\bibinfo  {journal} {Rev. Mod. Phys.}\ }\textbf {\bibinfo {volume} {92}},\
  \bibinfo {pages} {015003} (\bibinfo {year} {2020})}\BibitemShut {NoStop}%
\bibitem [{\citenamefont {Farhi}\ \emph {et~al.}(2000)\citenamefont {Farhi},
  \citenamefont {Goldstone}, \citenamefont {Gutmann},\ and\ \citenamefont
  {Sipser}}]{farhi2000adiabatic}%
  \BibitemOpen
  \bibfield  {author} {\bibinfo {author} {\bibfnamefont {E.}~\bibnamefont
  {Farhi}}, \bibinfo {author} {\bibfnamefont {J.}~\bibnamefont {Goldstone}},
  \bibinfo {author} {\bibfnamefont {S.}~\bibnamefont {Gutmann}},\ and\ \bibinfo
  {author} {\bibfnamefont {M.}~\bibnamefont {Sipser}},\ }\bibfield  {title}
  {\bibinfo {title} {Quantum computation by adiabatic evolution},\ }\href
  {https://arxiv.org/abs/quant-ph/0001106} {\bibfield  {journal} {\bibinfo
  {journal} {arXiv preprint arXiv:quant-ph/0001106}\ } (\bibinfo {year}
  {2000})}\BibitemShut {NoStop}%
\bibitem [{\citenamefont {Greiner}\ \emph {et~al.}(2002)\citenamefont
  {Greiner}, \citenamefont {Mandel}, \citenamefont {Esslinger}, \citenamefont
  {H{\"a}nsch},\ and\ \citenamefont {Bloch}}]{Greiner2002}%
  \BibitemOpen
  \bibfield  {author} {\bibinfo {author} {\bibfnamefont {M.}~\bibnamefont
  {Greiner}}, \bibinfo {author} {\bibfnamefont {O.}~\bibnamefont {Mandel}},
  \bibinfo {author} {\bibfnamefont {T.}~\bibnamefont {Esslinger}}, \bibinfo
  {author} {\bibfnamefont {T.~W.}\ \bibnamefont {H{\"a}nsch}},\ and\ \bibinfo
  {author} {\bibfnamefont {I.}~\bibnamefont {Bloch}},\ }\bibfield  {title}
  {\bibinfo {title} {Quantum phase transition from a superfluid to a mott
  insulator in a gas of ultracold atoms},\ }\href
  {https://doi.org/10.1038/415039a} {\bibfield  {journal} {\bibinfo  {journal}
  {Nature}\ }\textbf {\bibinfo {volume} {415}},\ \bibinfo {pages} {39}
  (\bibinfo {year} {2002})}\BibitemShut {NoStop}%
\bibitem [{\citenamefont {Albash}\ and\ \citenamefont
  {Lidar}(2018)}]{Albash2018adiabatic}%
  \BibitemOpen
  \bibfield  {author} {\bibinfo {author} {\bibfnamefont {T.}~\bibnamefont
  {Albash}}\ and\ \bibinfo {author} {\bibfnamefont {D.~A.}\ \bibnamefont
  {Lidar}},\ }\bibfield  {title} {\bibinfo {title} {Adiabatic quantum
  computation},\ }\href {https://doi.org/10.1103/RevModPhys.90.015002}
  {\bibfield  {journal} {\bibinfo  {journal} {Rev. Mod. Phys.}\ }\textbf
  {\bibinfo {volume} {90}},\ \bibinfo {pages} {015002} (\bibinfo {year}
  {2018})}\BibitemShut {NoStop}%
\bibitem [{\citenamefont {Claeys}\ \emph {et~al.}(2019)\citenamefont {Claeys},
  \citenamefont {Pandey}, \citenamefont {Sels},\ and\ \citenamefont
  {Polkovnikov}}]{claeys2019floquet}%
  \BibitemOpen
  \bibfield  {author} {\bibinfo {author} {\bibfnamefont {P.~W.}\ \bibnamefont
  {Claeys}}, \bibinfo {author} {\bibfnamefont {M.}~\bibnamefont {Pandey}},
  \bibinfo {author} {\bibfnamefont {D.}~\bibnamefont {Sels}},\ and\ \bibinfo
  {author} {\bibfnamefont {A.}~\bibnamefont {Polkovnikov}},\ }\bibfield
  {title} {\bibinfo {title} {Floquet-engineering counterdiabatic protocols in
  quantum many-body systems},\ }\href
  {https://doi.org/10.1103/PhysRevLett.123.090602} {\bibfield  {journal}
  {\bibinfo  {journal} {Phys. Rev. Lett.}\ }\textbf {\bibinfo {volume} {123}},\
  \bibinfo {pages} {090602} (\bibinfo {year} {2019})}\BibitemShut {NoStop}%
\bibitem [{\citenamefont {Semeghini}\ \emph {et~al.}(2021)\citenamefont
  {Semeghini}, \citenamefont {Levine}, \citenamefont {Keesling}, \citenamefont
  {Ebadi}, \citenamefont {Wang}, \citenamefont {Bluvstein}, \citenamefont
  {Verresen}, \citenamefont {Pichler}, \citenamefont {Kalinowski},
  \citenamefont {Samajdar}, \citenamefont {Omran}, \citenamefont {Sachdev},
  \citenamefont {Vishwanath}, \citenamefont {Greiner}, \citenamefont
  {Vuletić},\ and\ \citenamefont {Lukin}}]{Lukin2021adiabatic}%
  \BibitemOpen
  \bibfield  {author} {\bibinfo {author} {\bibfnamefont {G.}~\bibnamefont
  {Semeghini}}, \bibinfo {author} {\bibfnamefont {H.}~\bibnamefont {Levine}},
  \bibinfo {author} {\bibfnamefont {A.}~\bibnamefont {Keesling}}, \bibinfo
  {author} {\bibfnamefont {S.}~\bibnamefont {Ebadi}}, \bibinfo {author}
  {\bibfnamefont {T.~T.}\ \bibnamefont {Wang}}, \bibinfo {author}
  {\bibfnamefont {D.}~\bibnamefont {Bluvstein}}, \bibinfo {author}
  {\bibfnamefont {R.}~\bibnamefont {Verresen}}, \bibinfo {author}
  {\bibfnamefont {H.}~\bibnamefont {Pichler}}, \bibinfo {author} {\bibfnamefont
  {M.}~\bibnamefont {Kalinowski}}, \bibinfo {author} {\bibfnamefont
  {R.}~\bibnamefont {Samajdar}}, \bibinfo {author} {\bibfnamefont
  {A.}~\bibnamefont {Omran}}, \bibinfo {author} {\bibfnamefont
  {S.}~\bibnamefont {Sachdev}}, \bibinfo {author} {\bibfnamefont
  {A.}~\bibnamefont {Vishwanath}}, \bibinfo {author} {\bibfnamefont
  {M.}~\bibnamefont {Greiner}}, \bibinfo {author} {\bibfnamefont
  {V.}~\bibnamefont {Vuletić}},\ and\ \bibinfo {author} {\bibfnamefont
  {M.~D.}\ \bibnamefont {Lukin}},\ }\bibfield  {title} {\bibinfo {title}
  {Probing topological spin liquids on a programmable quantum simulator},\
  }\href {https://doi.org/10.1126/science.abi8794} {\bibfield  {journal}
  {\bibinfo  {journal} {Science}\ }\textbf {\bibinfo {volume} {374}},\ \bibinfo
  {pages} {1242} (\bibinfo {year} {2021})}\BibitemShut {NoStop}%
\bibitem [{\citenamefont {Cerezo}\ \emph {et~al.}(2021)\citenamefont {Cerezo},
  \citenamefont {Arrasmith}, \citenamefont {Babbush} \emph
  {et~al.}}]{Cerezo2021}%
  \BibitemOpen
  \bibfield  {author} {\bibinfo {author} {\bibfnamefont {M.}~\bibnamefont
  {Cerezo}}, \bibinfo {author} {\bibfnamefont {A.}~\bibnamefont {Arrasmith}},
  \bibinfo {author} {\bibfnamefont {R.}~\bibnamefont {Babbush}}, \emph
  {et~al.},\ }\bibfield  {title} {\bibinfo {title} {Variational quantum
  algorithms},\ }\href {https://doi.org/10.1038/s42254-021-00348-9} {\bibfield
  {journal} {\bibinfo  {journal} {Nature Reviews Physics}\ }\textbf {\bibinfo
  {volume} {3}},\ \bibinfo {pages} {625} (\bibinfo {year} {2021})}\BibitemShut
  {NoStop}%
\bibitem [{\citenamefont {Tilly}\ \emph {et~al.}(2022)\citenamefont {Tilly},
  \citenamefont {Chen}, \citenamefont {Cao}, \citenamefont {Picozzi},
  \citenamefont {Setia}, \citenamefont {Li}, \citenamefont {Grant},
  \citenamefont {Wossnig}, \citenamefont {Rungger}, \citenamefont {Booth},\
  and\ \citenamefont {Tennyson}}]{Tilly2022variational}%
  \BibitemOpen
  \bibfield  {author} {\bibinfo {author} {\bibfnamefont {J.}~\bibnamefont
  {Tilly}}, \bibinfo {author} {\bibfnamefont {H.}~\bibnamefont {Chen}},
  \bibinfo {author} {\bibfnamefont {S.}~\bibnamefont {Cao}}, \bibinfo {author}
  {\bibfnamefont {D.}~\bibnamefont {Picozzi}}, \bibinfo {author} {\bibfnamefont
  {K.}~\bibnamefont {Setia}}, \bibinfo {author} {\bibfnamefont
  {Y.}~\bibnamefont {Li}}, \bibinfo {author} {\bibfnamefont {E.}~\bibnamefont
  {Grant}}, \bibinfo {author} {\bibfnamefont {L.}~\bibnamefont {Wossnig}},
  \bibinfo {author} {\bibfnamefont {I.}~\bibnamefont {Rungger}}, \bibinfo
  {author} {\bibfnamefont {G.~H.}\ \bibnamefont {Booth}},\ and\ \bibinfo
  {author} {\bibfnamefont {J.}~\bibnamefont {Tennyson}},\ }\bibfield  {title}
  {\bibinfo {title} {The variational quantum eigensolver: A review of methods
  and best practices},\ }\href
  {https://doi.org/https://doi.org/10.1016/j.physrep.2022.08.003} {\bibfield
  {journal} {\bibinfo  {journal} {Physics Reports}\ }\textbf {\bibinfo {volume}
  {986}},\ \bibinfo {pages} {1} (\bibinfo {year} {2022})}\BibitemShut {NoStop}%
\bibitem [{\citenamefont {Fedorov}\ \emph {et~al.}(2022)\citenamefont
  {Fedorov}, \citenamefont {Peng}, \citenamefont {Govind}, \citenamefont
  {Zhang},\ and\ \citenamefont {et~al.}}]{Fedorov2022variational}%
  \BibitemOpen
  \bibfield  {author} {\bibinfo {author} {\bibfnamefont {D.~A.}\ \bibnamefont
  {Fedorov}}, \bibinfo {author} {\bibfnamefont {B.}~\bibnamefont {Peng}},
  \bibinfo {author} {\bibfnamefont {N.}~\bibnamefont {Govind}}, \bibinfo
  {author} {\bibfnamefont {H.}~\bibnamefont {Zhang}},\ and\ \bibinfo {author}
  {\bibnamefont {et~al.}},\ }\bibfield  {title} {\bibinfo {title} {Vqe method:
  a short survey and recent developments},\ }\href
  {https://doi.org/10.1186/s41313-021-00032-6} {\bibfield  {journal} {\bibinfo
  {journal} {Materials Theory}\ }\textbf {\bibinfo {volume} {6}},\ \bibinfo
  {pages} {2} (\bibinfo {year} {2022})}\BibitemShut {NoStop}%
\bibitem [{\citenamefont {Poyatos}\ \emph {et~al.}(1996)\citenamefont
  {Poyatos}, \citenamefont {Cirac},\ and\ \citenamefont {Zoller}}]{Zoller1996}%
  \BibitemOpen
  \bibfield  {author} {\bibinfo {author} {\bibfnamefont {J.~F.}\ \bibnamefont
  {Poyatos}}, \bibinfo {author} {\bibfnamefont {J.~I.}\ \bibnamefont {Cirac}},\
  and\ \bibinfo {author} {\bibfnamefont {P.}~\bibnamefont {Zoller}},\
  }\bibfield  {title} {\bibinfo {title} {Quantum reservoir engineering with
  laser cooled trapped ions},\ }\href
  {https://doi.org/10.1103/PhysRevLett.77.4728} {\bibfield  {journal} {\bibinfo
   {journal} {Phys. Rev. Lett.}\ }\textbf {\bibinfo {volume} {77}},\ \bibinfo
  {pages} {4728} (\bibinfo {year} {1996})}\BibitemShut {NoStop}%
\bibitem [{\citenamefont {Pielawa}\ \emph {et~al.}(2007)\citenamefont
  {Pielawa}, \citenamefont {Morigi}, \citenamefont {Vitali},\ and\
  \citenamefont {Davidovich}}]{Giovanna2007}%
  \BibitemOpen
  \bibfield  {author} {\bibinfo {author} {\bibfnamefont {S.}~\bibnamefont
  {Pielawa}}, \bibinfo {author} {\bibfnamefont {G.}~\bibnamefont {Morigi}},
  \bibinfo {author} {\bibfnamefont {D.}~\bibnamefont {Vitali}},\ and\ \bibinfo
  {author} {\bibfnamefont {L.}~\bibnamefont {Davidovich}},\ }\bibfield  {title}
  {\bibinfo {title} {Generation of einstein-podolsky-rosen-entangled radiation
  through an atomic reservoir},\ }\href
  {https://doi.org/10.1103/PhysRevLett.98.240401} {\bibfield  {journal}
  {\bibinfo  {journal} {Phys. Rev. Lett.}\ }\textbf {\bibinfo {volume} {98}},\
  \bibinfo {pages} {240401} (\bibinfo {year} {2007})}\BibitemShut {NoStop}%
\bibitem [{\citenamefont {Diehl}\ \emph {et~al.}(2008)\citenamefont {Diehl},
  \citenamefont {Micheli}, \citenamefont {Kantian}, \citenamefont {Kraus},
  \citenamefont {Büchler},\ and\ \citenamefont {Zoller}}]{Diehl2008}%
  \BibitemOpen
  \bibfield  {author} {\bibinfo {author} {\bibfnamefont {S.}~\bibnamefont
  {Diehl}}, \bibinfo {author} {\bibfnamefont {A.}~\bibnamefont {Micheli}},
  \bibinfo {author} {\bibfnamefont {A.}~\bibnamefont {Kantian}}, \bibinfo
  {author} {\bibfnamefont {B.}~\bibnamefont {Kraus}}, \bibinfo {author}
  {\bibfnamefont {H.~P.}\ \bibnamefont {Büchler}},\ and\ \bibinfo {author}
  {\bibfnamefont {P.}~\bibnamefont {Zoller}},\ }\bibfield  {title} {\bibinfo
  {title} {Quantum states and phases in driven open quantum systems with cold
  atoms},\ }\href {https://doi.org/10.1038/nphys1073} {\bibfield  {journal}
  {\bibinfo  {journal} {Nature Physics}\ }\textbf {\bibinfo {volume} {4}},\
  \bibinfo {pages} {878} (\bibinfo {year} {2008})}\BibitemShut {NoStop}%
\bibitem [{\citenamefont {Verstraete}\ \emph {et~al.}(2009)\citenamefont
  {Verstraete}, \citenamefont {Wolf},\ and\ \citenamefont
  {Cirac}}]{Verstraete2009}%
  \BibitemOpen
  \bibfield  {author} {\bibinfo {author} {\bibfnamefont {F.}~\bibnamefont
  {Verstraete}}, \bibinfo {author} {\bibfnamefont {M.~M.}\ \bibnamefont
  {Wolf}},\ and\ \bibinfo {author} {\bibfnamefont {J.~I.}\ \bibnamefont
  {Cirac}},\ }\bibfield  {title} {\bibinfo {title} {Quantum computation and
  quantum-state engineering driven by dissipation},\ }\href
  {https://doi.org/10.1038/nphys1342} {\bibfield  {journal} {\bibinfo
  {journal} {Nature Physics}\ }\textbf {\bibinfo {volume} {5}},\ \bibinfo
  {pages} {633} (\bibinfo {year} {2009})}\BibitemShut {NoStop}%
\bibitem [{\citenamefont {Schirmer}\ and\ \citenamefont
  {Wang}(2010)}]{Xiaoting2010}%
  \BibitemOpen
  \bibfield  {author} {\bibinfo {author} {\bibfnamefont {S.~G.}\ \bibnamefont
  {Schirmer}}\ and\ \bibinfo {author} {\bibfnamefont {X.}~\bibnamefont
  {Wang}},\ }\bibfield  {title} {\bibinfo {title} {Stabilizing open quantum
  systems by markovian reservoir engineering},\ }\href
  {https://doi.org/10.1103/PhysRevA.81.062306} {\bibfield  {journal} {\bibinfo
  {journal} {Phys. Rev. A}\ }\textbf {\bibinfo {volume} {81}},\ \bibinfo
  {pages} {062306} (\bibinfo {year} {2010})}\BibitemShut {NoStop}%
\bibitem [{\citenamefont {Muschik}\ \emph {et~al.}(2011)\citenamefont
  {Muschik}, \citenamefont {Polzik},\ and\ \citenamefont {Cirac}}]{Cirac2011a}%
  \BibitemOpen
  \bibfield  {author} {\bibinfo {author} {\bibfnamefont {C.~A.}\ \bibnamefont
  {Muschik}}, \bibinfo {author} {\bibfnamefont {E.~S.}\ \bibnamefont
  {Polzik}},\ and\ \bibinfo {author} {\bibfnamefont {J.~I.}\ \bibnamefont
  {Cirac}},\ }\bibfield  {title} {\bibinfo {title} {Dissipatively driven
  entanglement of two macroscopic atomic ensembles},\ }\href
  {https://doi.org/10.1103/PhysRevA.83.052312} {\bibfield  {journal} {\bibinfo
  {journal} {Phys. Rev. A}\ }\textbf {\bibinfo {volume} {83}},\ \bibinfo
  {pages} {052312} (\bibinfo {year} {2011})}\BibitemShut {NoStop}%
\bibitem [{\citenamefont {Barreiro}\ \emph {et~al.}(2011)\citenamefont
  {Barreiro}, \citenamefont {Müller}, \citenamefont {Schindler}, \citenamefont
  {Nigg}, \citenamefont {Monz}, \citenamefont {Chwalla}, \citenamefont
  {Hennrich}, \citenamefont {Roos}, \citenamefont {Zoller},\ and\ \citenamefont
  {Blatt}}]{Barreiro2011}%
  \BibitemOpen
  \bibfield  {author} {\bibinfo {author} {\bibfnamefont {J.}~\bibnamefont
  {Barreiro}}, \bibinfo {author} {\bibfnamefont {M.}~\bibnamefont {Müller}},
  \bibinfo {author} {\bibfnamefont {P.}~\bibnamefont {Schindler}}, \bibinfo
  {author} {\bibfnamefont {D.}~\bibnamefont {Nigg}}, \bibinfo {author}
  {\bibfnamefont {T.}~\bibnamefont {Monz}}, \bibinfo {author} {\bibfnamefont
  {M.}~\bibnamefont {Chwalla}}, \bibinfo {author} {\bibfnamefont
  {M.}~\bibnamefont {Hennrich}}, \bibinfo {author} {\bibfnamefont
  {C.}~\bibnamefont {Roos}}, \bibinfo {author} {\bibfnamefont {P.}~\bibnamefont
  {Zoller}},\ and\ \bibinfo {author} {\bibfnamefont {R.}~\bibnamefont
  {Blatt}},\ }\bibfield  {title} {\bibinfo {title} {An open-system quantum
  simulator with trapped ions},\ }\href {https://doi.org/10.1038/nature09801}
  {\bibfield  {journal} {\bibinfo  {journal} {Nature}\ }\textbf {\bibinfo
  {volume} {470}},\ \bibinfo {pages} {486} (\bibinfo {year}
  {2011})}\BibitemShut {NoStop}%
\bibitem [{\citenamefont {Shankar}\ \emph {et~al.}(2013)\citenamefont
  {Shankar}, \citenamefont {Hatridge}, \citenamefont {Leghtas}, \citenamefont
  {Sliwa}, \citenamefont {Narla}, \citenamefont {Vool}, \citenamefont {Girvin},
  \citenamefont {Frunzio}, \citenamefont {Schoelkopf},\ and\ \citenamefont
  {Devoret}}]{Shankar2013}%
  \BibitemOpen
  \bibfield  {author} {\bibinfo {author} {\bibfnamefont {S.}~\bibnamefont
  {Shankar}}, \bibinfo {author} {\bibfnamefont {M.}~\bibnamefont {Hatridge}},
  \bibinfo {author} {\bibfnamefont {Z.}~\bibnamefont {Leghtas}}, \bibinfo
  {author} {\bibfnamefont {K.~M.}\ \bibnamefont {Sliwa}}, \bibinfo {author}
  {\bibfnamefont {A.}~\bibnamefont {Narla}}, \bibinfo {author} {\bibfnamefont
  {U.}~\bibnamefont {Vool}}, \bibinfo {author} {\bibfnamefont {S.~M.}\
  \bibnamefont {Girvin}}, \bibinfo {author} {\bibfnamefont {L.}~\bibnamefont
  {Frunzio}}, \bibinfo {author} {\bibfnamefont {R.~J.}\ \bibnamefont
  {Schoelkopf}},\ and\ \bibinfo {author} {\bibfnamefont {M.~H.}\ \bibnamefont
  {Devoret}},\ }\bibfield  {title} {\bibinfo {title} {Autonomously stabilized
  entanglement between two superconducting quantum bits},\ }\href
  {https://doi.org/10.1038/nature12802} {\bibfield  {journal} {\bibinfo
  {journal} {Nature}\ }\textbf {\bibinfo {volume} {504}},\ \bibinfo {pages}
  {419} (\bibinfo {year} {2013})}\BibitemShut {NoStop}%
\bibitem [{\citenamefont {Lin}\ \emph {et~al.}(2013)\citenamefont {Lin},
  \citenamefont {Gaebler}, \citenamefont {Reiter}, \citenamefont {Tan},
  \citenamefont {Bowler}, \citenamefont {Soerensen}, \citenamefont
  {Leibfried},\ and\ \citenamefont {Wineland}}]{Lin2013}%
  \BibitemOpen
  \bibfield  {author} {\bibinfo {author} {\bibfnamefont {Y.}~\bibnamefont
  {Lin}}, \bibinfo {author} {\bibfnamefont {J.~P.}\ \bibnamefont {Gaebler}},
  \bibinfo {author} {\bibfnamefont {F.}~\bibnamefont {Reiter}}, \bibinfo
  {author} {\bibfnamefont {T.~R.}\ \bibnamefont {Tan}}, \bibinfo {author}
  {\bibfnamefont {R.}~\bibnamefont {Bowler}}, \bibinfo {author} {\bibfnamefont
  {A.~S.}\ \bibnamefont {Soerensen}}, \bibinfo {author} {\bibfnamefont
  {D.}~\bibnamefont {Leibfried}},\ and\ \bibinfo {author} {\bibfnamefont
  {D.~J.}\ \bibnamefont {Wineland}},\ }\bibfield  {title} {\bibinfo {title}
  {Dissipative production of a maximally entangled steady state of two quantum
  bits},\ }\href {https://doi.org/10.1038/nature12801} {\bibfield  {journal}
  {\bibinfo  {journal} {Nature}\ }\textbf {\bibinfo {volume} {504}},\ \bibinfo
  {pages} {415} (\bibinfo {year} {2013})}\BibitemShut {NoStop}%
\bibitem [{\citenamefont {Morigi}\ \emph {et~al.}(2015)\citenamefont {Morigi},
  \citenamefont {Eschner}, \citenamefont {Cormick}, \citenamefont {Lin},
  \citenamefont {Leibfried},\ and\ \citenamefont {Wineland}}]{Giovanna2015}%
  \BibitemOpen
  \bibfield  {author} {\bibinfo {author} {\bibfnamefont {G.}~\bibnamefont
  {Morigi}}, \bibinfo {author} {\bibfnamefont {J.}~\bibnamefont {Eschner}},
  \bibinfo {author} {\bibfnamefont {C.}~\bibnamefont {Cormick}}, \bibinfo
  {author} {\bibfnamefont {Y.}~\bibnamefont {Lin}}, \bibinfo {author}
  {\bibfnamefont {D.}~\bibnamefont {Leibfried}},\ and\ \bibinfo {author}
  {\bibfnamefont {D.~J.}\ \bibnamefont {Wineland}},\ }\bibfield  {title}
  {\bibinfo {title} {Dissipative quantum control of a spin chain},\ }\href
  {https://doi.org/10.1103/PhysRevLett.115.200502} {\bibfield  {journal}
  {\bibinfo  {journal} {Phys. Rev. Lett.}\ }\textbf {\bibinfo {volume} {115}},\
  \bibinfo {pages} {200502} (\bibinfo {year} {2015})}\BibitemShut {NoStop}%
\bibitem [{\citenamefont {Reiter}\ \emph {et~al.}(2016)\citenamefont {Reiter},
  \citenamefont {Reeb},\ and\ \citenamefont {S\o{}rensen}}]{Reiter2016}%
  \BibitemOpen
  \bibfield  {author} {\bibinfo {author} {\bibfnamefont {F.}~\bibnamefont
  {Reiter}}, \bibinfo {author} {\bibfnamefont {D.}~\bibnamefont {Reeb}},\ and\
  \bibinfo {author} {\bibfnamefont {A.~S.}\ \bibnamefont {S\o{}rensen}},\
  }\bibfield  {title} {\bibinfo {title} {Scalable dissipative preparation of
  many-body entanglement},\ }\href
  {https://doi.org/10.1103/PhysRevLett.117.040501} {\bibfield  {journal}
  {\bibinfo  {journal} {Phys. Rev. Lett.}\ }\textbf {\bibinfo {volume} {117}},\
  \bibinfo {pages} {040501} (\bibinfo {year} {2016})}\BibitemShut {NoStop}%
\bibitem [{\citenamefont {Motzoi}\ \emph {et~al.}(2016)\citenamefont {Motzoi},
  \citenamefont {Halperin}, \citenamefont {Wang}, \citenamefont {Whaley},\ and\
  \citenamefont {Schirmer}}]{Xiaoting2016}%
  \BibitemOpen
  \bibfield  {author} {\bibinfo {author} {\bibfnamefont {F.}~\bibnamefont
  {Motzoi}}, \bibinfo {author} {\bibfnamefont {E.}~\bibnamefont {Halperin}},
  \bibinfo {author} {\bibfnamefont {X.}~\bibnamefont {Wang}}, \bibinfo {author}
  {\bibfnamefont {K.~B.}\ \bibnamefont {Whaley}},\ and\ \bibinfo {author}
  {\bibfnamefont {S.}~\bibnamefont {Schirmer}},\ }\bibfield  {title} {\bibinfo
  {title} {Backaction-driven, robust, steady-state long-distance qubit
  entanglement over lossy channels},\ }\href
  {https://doi.org/10.1103/PhysRevA.94.032313} {\bibfield  {journal} {\bibinfo
  {journal} {Phys. Rev. A}\ }\textbf {\bibinfo {volume} {94}},\ \bibinfo
  {pages} {032313} (\bibinfo {year} {2016})}\BibitemShut {NoStop}%
\bibitem [{\citenamefont {Roy}\ \emph {et~al.}(2020)\citenamefont {Roy},
  \citenamefont {Chalker}, \citenamefont {Gornyi},\ and\ \citenamefont
  {Gefen}}]{Yuval2020Steer}%
  \BibitemOpen
  \bibfield  {author} {\bibinfo {author} {\bibfnamefont {S.}~\bibnamefont
  {Roy}}, \bibinfo {author} {\bibfnamefont {J.~T.}\ \bibnamefont {Chalker}},
  \bibinfo {author} {\bibfnamefont {I.~V.}\ \bibnamefont {Gornyi}},\ and\
  \bibinfo {author} {\bibfnamefont {Y.}~\bibnamefont {Gefen}},\ }\bibfield
  {title} {\bibinfo {title} {Measurement-induced steering of quantum systems},\
  }\href {https://doi.org/10.1103/PhysRevResearch.2.033347} {\bibfield
  {journal} {\bibinfo  {journal} {Phys. Rev. Res.}\ }\textbf {\bibinfo {volume}
  {2}},\ \bibinfo {pages} {033347} (\bibinfo {year} {2020})}\BibitemShut
  {NoStop}%
\bibitem [{\citenamefont {Zhou}\ \emph {et~al.}(2021)\citenamefont {Zhou},
  \citenamefont {Choi},\ and\ \citenamefont {Lukin}}]{Soonwon2021}%
  \BibitemOpen
  \bibfield  {author} {\bibinfo {author} {\bibfnamefont {L.}~\bibnamefont
  {Zhou}}, \bibinfo {author} {\bibfnamefont {S.}~\bibnamefont {Choi}},\ and\
  \bibinfo {author} {\bibfnamefont {M.~D.}\ \bibnamefont {Lukin}},\ }\bibfield
  {title} {\bibinfo {title} {Symmetry-protected dissipative preparation of
  matrix product states},\ }\href {https://doi.org/10.1103/PhysRevA.104.032418}
  {\bibfield  {journal} {\bibinfo  {journal} {Phys. Rev. A}\ }\textbf {\bibinfo
  {volume} {104}},\ \bibinfo {pages} {032418} (\bibinfo {year}
  {2021})}\BibitemShut {NoStop}%
\bibitem [{\citenamefont {Rall}\ \emph {et~al.}(2023)\citenamefont {Rall},
  \citenamefont {Wang},\ and\ \citenamefont {Wocjan}}]{Rall2023thermalstate}%
  \BibitemOpen
  \bibfield  {author} {\bibinfo {author} {\bibfnamefont {P.}~\bibnamefont
  {Rall}}, \bibinfo {author} {\bibfnamefont {C.}~\bibnamefont {Wang}},\ and\
  \bibinfo {author} {\bibfnamefont {P.}~\bibnamefont {Wocjan}},\ }\bibfield
  {title} {\bibinfo {title} {Thermal {S}tate {P}reparation via {R}ounding
  {P}romises},\ }\href {https://doi.org/10.22331/q-2023-10-10-1132} {\bibfield
  {journal} {\bibinfo  {journal} {{Quantum}}\ }\textbf {\bibinfo {volume}
  {7}},\ \bibinfo {pages} {1132} (\bibinfo {year} {2023})}\BibitemShut
  {NoStop}%
\bibitem [{\citenamefont {Chen}\ \emph {et~al.}(2023)\citenamefont {Chen},
  \citenamefont {Kastoryano}, \citenamefont {Brandão},\ and\ \citenamefont
  {Gilyén}}]{chen2023thermal}%
  \BibitemOpen
  \bibfield  {author} {\bibinfo {author} {\bibfnamefont {C.-F.}\ \bibnamefont
  {Chen}}, \bibinfo {author} {\bibfnamefont {M.~J.}\ \bibnamefont
  {Kastoryano}}, \bibinfo {author} {\bibfnamefont {F.~G. S.~L.}\ \bibnamefont
  {Brandão}},\ and\ \bibinfo {author} {\bibfnamefont {A.}~\bibnamefont
  {Gilyén}},\ }\bibfield  {title} {\bibinfo {title} {Quantum thermal state
  preparation},\ }\href {https://arxiv.org/abs/2303.18224} {\bibfield
  {journal} {\bibinfo  {journal} {arXiv preprint arXiv:2303.18224}\ } (\bibinfo
  {year} {2023})}\BibitemShut {NoStop}%
\bibitem [{\citenamefont {Mori}(2023)}]{Takashi2023thermal}%
  \BibitemOpen
  \bibfield  {author} {\bibinfo {author} {\bibfnamefont {T.}~\bibnamefont
  {Mori}},\ }\bibfield  {title} {\bibinfo {title} {Floquet states in open
  quantum systems},\ }\href
  {https://doi.org/https://doi.org/10.1146/annurev-conmatphys-040721-015537}
  {\bibfield  {journal} {\bibinfo  {journal} {Annual Review of Condensed Matter
  Physics}\ }\textbf {\bibinfo {volume} {14}},\ \bibinfo {pages} {35} (\bibinfo
  {year} {2023})}\BibitemShut {NoStop}%
\bibitem [{\citenamefont {Shtanko}\ and\ \citenamefont
  {Movassagh}(2021)}]{shtanko2021thermal}%
  \BibitemOpen
  \bibfield  {author} {\bibinfo {author} {\bibfnamefont {O.}~\bibnamefont
  {Shtanko}}\ and\ \bibinfo {author} {\bibfnamefont {R.}~\bibnamefont
  {Movassagh}},\ }\bibfield  {title} {\bibinfo {title} {Preparing thermal
  states on noiseless and noisy programmable quantum processors},\ }\href
  {https://doi.org/10.48550/arXiv.2112.14688} {\bibfield  {journal} {\bibinfo
  {journal} {arXiv preprint arXiv:2112.14688}\ } (\bibinfo {year}
  {2021})}\BibitemShut {NoStop}%
\bibitem [{\citenamefont {Puente}\ \emph {et~al.}(2024)\citenamefont {Puente},
  \citenamefont {Motzoi}, \citenamefont {Calarco}, \citenamefont {Morigi},\
  and\ \citenamefont {Rizzi}}]{Giovanna2024quantumstate}%
  \BibitemOpen
  \bibfield  {author} {\bibinfo {author} {\bibfnamefont {D.~A.}\ \bibnamefont
  {Puente}}, \bibinfo {author} {\bibfnamefont {F.}~\bibnamefont {Motzoi}},
  \bibinfo {author} {\bibfnamefont {T.}~\bibnamefont {Calarco}}, \bibinfo
  {author} {\bibfnamefont {G.}~\bibnamefont {Morigi}},\ and\ \bibinfo {author}
  {\bibfnamefont {M.}~\bibnamefont {Rizzi}},\ }\bibfield  {title} {\bibinfo
  {title} {Quantum state preparation via engineered ancilla resetting},\ }\href
  {https://doi.org/10.22331/q-2024-03-27-1299} {\bibfield  {journal} {\bibinfo
  {journal} {{Quantum}}\ }\textbf {\bibinfo {volume} {8}},\ \bibinfo {pages}
  {1299} (\bibinfo {year} {2024})}\BibitemShut {NoStop}%
\bibitem [{\citenamefont {Matthies}\ \emph {et~al.}(2024)\citenamefont
  {Matthies}, \citenamefont {Rudner}, \citenamefont {Rosch},\ and\
  \citenamefont {Berg}}]{Matthies2024programmable}%
  \BibitemOpen
  \bibfield  {author} {\bibinfo {author} {\bibfnamefont {A.}~\bibnamefont
  {Matthies}}, \bibinfo {author} {\bibfnamefont {M.}~\bibnamefont {Rudner}},
  \bibinfo {author} {\bibfnamefont {A.}~\bibnamefont {Rosch}},\ and\ \bibinfo
  {author} {\bibfnamefont {E.}~\bibnamefont {Berg}},\ }\bibfield  {title}
  {\bibinfo {title} {Programmable adiabatic demagnetization for systems with
  trivial and topological excitations},\ }\href
  {https://doi.org/10.22331/q-2024-10-23-1505} {\bibfield  {journal} {\bibinfo
  {journal} {{Quantum}}\ }\textbf {\bibinfo {volume} {8}},\ \bibinfo {pages}
  {1505} (\bibinfo {year} {2024})}\BibitemShut {NoStop}%
\bibitem [{\citenamefont {Mi}\ \emph {et~al.}(2024)\citenamefont {Mi} \emph
  {et~al.}}]{Xiao2024Google}%
  \BibitemOpen
  \bibfield  {author} {\bibinfo {author} {\bibfnamefont {X.}~\bibnamefont {Mi}}
  \emph {et~al.},\ }\bibfield  {title} {\bibinfo {title} {Stable
  quantum-correlated many-body states through engineered dissipation},\ }\href
  {https://doi.org/10.1126/science.adh9932} {\bibfield  {journal} {\bibinfo
  {journal} {Science}\ }\textbf {\bibinfo {volume} {383}},\ \bibinfo {pages}
  {1332} (\bibinfo {year} {2024})}\BibitemShut {NoStop}%
\bibitem [{\citenamefont {Lloyd}\ \emph {et~al.}(2025)\citenamefont {Lloyd},
  \citenamefont {Michailidis}, \citenamefont {Mi}, \citenamefont
  {Smelyanskiy},\ and\ \citenamefont {Abanin}}]{Xiao2025}%
  \BibitemOpen
  \bibfield  {author} {\bibinfo {author} {\bibfnamefont {J.}~\bibnamefont
  {Lloyd}}, \bibinfo {author} {\bibfnamefont {A.~A.}\ \bibnamefont
  {Michailidis}}, \bibinfo {author} {\bibfnamefont {X.}~\bibnamefont {Mi}},
  \bibinfo {author} {\bibfnamefont {V.}~\bibnamefont {Smelyanskiy}},\ and\
  \bibinfo {author} {\bibfnamefont {D.~A.}\ \bibnamefont {Abanin}},\ }\bibfield
   {title} {\bibinfo {title} {Quasiparticle cooling algorithms for quantum
  many-body state preparation},\ }\href
  {https://doi.org/10.1103/PRXQuantum.6.010361} {\bibfield  {journal} {\bibinfo
   {journal} {PRX Quantum}\ }\textbf {\bibinfo {volume} {6}},\ \bibinfo {pages}
  {010361} (\bibinfo {year} {2025})}\BibitemShut {NoStop}%
\bibitem [{\citenamefont {Kraus}\ \emph {et~al.}(2008)\citenamefont {Kraus},
  \citenamefont {B\"uchler}, \citenamefont {Diehl}, \citenamefont {Kantian},
  \citenamefont {Micheli},\ and\ \citenamefont {Zoller}}]{Zoller2008pra}%
  \BibitemOpen
  \bibfield  {author} {\bibinfo {author} {\bibfnamefont {B.}~\bibnamefont
  {Kraus}}, \bibinfo {author} {\bibfnamefont {H.~P.}\ \bibnamefont
  {B\"uchler}}, \bibinfo {author} {\bibfnamefont {S.}~\bibnamefont {Diehl}},
  \bibinfo {author} {\bibfnamefont {A.}~\bibnamefont {Kantian}}, \bibinfo
  {author} {\bibfnamefont {A.}~\bibnamefont {Micheli}},\ and\ \bibinfo {author}
  {\bibfnamefont {P.}~\bibnamefont {Zoller}},\ }\bibfield  {title} {\bibinfo
  {title} {Preparation of entangled states by quantum markov processes},\
  }\href {https://doi.org/10.1103/PhysRevA.78.042307} {\bibfield  {journal}
  {\bibinfo  {journal} {Phys. Rev. A}\ }\textbf {\bibinfo {volume} {78}},\
  \bibinfo {pages} {042307} (\bibinfo {year} {2008})}\BibitemShut {NoStop}%
\bibitem [{\citenamefont {Lu}\ \emph {et~al.}(2023)\citenamefont {Lu},
  \citenamefont {Zhang}, \citenamefont {Vijay},\ and\ \citenamefont
  {Hsieh}}]{lu2023mixed}%
  \BibitemOpen
  \bibfield  {author} {\bibinfo {author} {\bibfnamefont {T.-C.}\ \bibnamefont
  {Lu}}, \bibinfo {author} {\bibfnamefont {Z.}~\bibnamefont {Zhang}}, \bibinfo
  {author} {\bibfnamefont {S.}~\bibnamefont {Vijay}},\ and\ \bibinfo {author}
  {\bibfnamefont {T.~H.}\ \bibnamefont {Hsieh}},\ }\bibfield  {title} {\bibinfo
  {title} {Mixed-state long-range order and criticality from measurement and
  feedback},\ }\href {https://doi.org/10.1103/PRXQuantum.4.030318} {\bibfield
  {journal} {\bibinfo  {journal} {PRX Quantum}\ }\textbf {\bibinfo {volume}
  {4}},\ \bibinfo {pages} {030318} (\bibinfo {year} {2023})}\BibitemShut
  {NoStop}%
\bibitem [{\citenamefont {Lavasani}\ \emph {et~al.}(2023)\citenamefont
  {Lavasani}, \citenamefont {Luo},\ and\ \citenamefont
  {Vijay}}]{lavasani2023monitored}%
  \BibitemOpen
  \bibfield  {author} {\bibinfo {author} {\bibfnamefont {A.}~\bibnamefont
  {Lavasani}}, \bibinfo {author} {\bibfnamefont {Z.-X.}\ \bibnamefont {Luo}},\
  and\ \bibinfo {author} {\bibfnamefont {S.}~\bibnamefont {Vijay}},\ }\bibfield
   {title} {\bibinfo {title} {Monitored quantum dynamics and the kitaev spin
  liquid},\ }\href {https://doi.org/10.1103/PhysRevB.108.115135} {\bibfield
  {journal} {\bibinfo  {journal} {Phys. Rev. B}\ }\textbf {\bibinfo {volume}
  {108}},\ \bibinfo {pages} {115135} (\bibinfo {year} {2023})}\BibitemShut
  {NoStop}%
\bibitem [{\citenamefont {Ritter}\ \emph {et~al.}(2025)\citenamefont {Ritter},
  \citenamefont {Long}, \citenamefont {Yue}, \citenamefont {Chandran},\ and\
  \citenamefont {Kollár}}]{ritter2024autonomous}%
  \BibitemOpen
  \bibfield  {author} {\bibinfo {author} {\bibfnamefont {M.}~\bibnamefont
  {Ritter}}, \bibinfo {author} {\bibfnamefont {D.~M.}\ \bibnamefont {Long}},
  \bibinfo {author} {\bibfnamefont {Q.}~\bibnamefont {Yue}}, \bibinfo {author}
  {\bibfnamefont {A.}~\bibnamefont {Chandran}},\ and\ \bibinfo {author}
  {\bibfnamefont {A.~J.}\ \bibnamefont {Kollár}},\ }\bibfield  {title}
  {\bibinfo {title} {Autonomous stabilization of floquet states using static
  dissipation},\ }\href {https://arxiv.org/abs/2410.12908} {\bibfield
  {journal} {\bibinfo  {journal} {arXiv preprint arXiv:2410.12908}\ } (\bibinfo
  {year} {2025})}\BibitemShut {NoStop}%
\bibitem [{\citenamefont {Schnell}\ \emph {et~al.}(2024)\citenamefont
  {Schnell}, \citenamefont {Weitenberg},\ and\ \citenamefont
  {Eckardt}}]{schnell2024dissipative}%
  \BibitemOpen
  \bibfield  {author} {\bibinfo {author} {\bibfnamefont {A.}~\bibnamefont
  {Schnell}}, \bibinfo {author} {\bibfnamefont {C.}~\bibnamefont
  {Weitenberg}},\ and\ \bibinfo {author} {\bibfnamefont {A.}~\bibnamefont
  {Eckardt}},\ }\bibfield  {title} {\bibinfo {title} {{Dissipative preparation
  of a Floquet topological insulator in an optical lattice via bath
  engineering}},\ }\href {https://doi.org/10.21468/SciPostPhys.17.2.052}
  {\bibfield  {journal} {\bibinfo  {journal} {SciPost Phys.}\ }\textbf
  {\bibinfo {volume} {17}},\ \bibinfo {pages} {052} (\bibinfo {year}
  {2024})}\BibitemShut {NoStop}%
\bibitem [{\citenamefont {Zhao}\ \emph {et~al.}(2025)\citenamefont {Zhao},
  \citenamefont {Wu}, \citenamefont {Petiziol},\ and\ \citenamefont
  {Eckardt}}]{zhao2025feedback}%
  \BibitemOpen
  \bibfield  {author} {\bibinfo {author} {\bibfnamefont {W.}~\bibnamefont
  {Zhao}}, \bibinfo {author} {\bibfnamefont {L.-N.}\ \bibnamefont {Wu}},
  \bibinfo {author} {\bibfnamefont {F.}~\bibnamefont {Petiziol}},\ and\
  \bibinfo {author} {\bibfnamefont {A.}~\bibnamefont {Eckardt}},\ }\bibfield
  {title} {\bibinfo {title} {Feedback cooling of fermionic atoms in optical
  lattices},\ }\href {https://doi.org/10.48550/arXiv.2501.07293} {\bibfield
  {journal} {\bibinfo  {journal} {arXiv preprint arXiv:2501.07293}\ } (\bibinfo
  {year} {2025})}\BibitemShut {NoStop}%
\bibitem [{\citenamefont {Lin}(2025)}]{lin2025dissipative}%
  \BibitemOpen
  \bibfield  {author} {\bibinfo {author} {\bibfnamefont {L.}~\bibnamefont
  {Lin}},\ }\bibfield  {title} {\bibinfo {title} {Dissipative preparation of
  many-body quantum states: Towards practical quantum advantage},\ }\href
  {https://arxiv.org/abs/2505.21308} {\bibfield  {journal} {\bibinfo  {journal}
  {arXiv preprint arXiv:2505.21308}\ } (\bibinfo {year} {2025})}\BibitemShut
  {NoStop}%
\bibitem [{\citenamefont {Bu{\v{c}}a}\ \emph {et~al.}(2019)\citenamefont
  {Bu{\v{c}}a}, \citenamefont {Tindall},\ and\ \citenamefont
  {Jaksch}}]{Buca2019}%
  \BibitemOpen
  \bibfield  {author} {\bibinfo {author} {\bibfnamefont {B.}~\bibnamefont
  {Bu{\v{c}}a}}, \bibinfo {author} {\bibfnamefont {J.}~\bibnamefont
  {Tindall}},\ and\ \bibinfo {author} {\bibfnamefont {D.}~\bibnamefont
  {Jaksch}},\ }\bibfield  {title} {\bibinfo {title} {Non‐stationary coherent
  quantum many‐body dynamics through dissipation},\ }\href
  {https://doi.org/10.1038/s41467-019-09757-y} {\bibfield  {journal} {\bibinfo
  {journal} {Nature Communications}\ }\textbf {\bibinfo {volume} {10}},\
  \bibinfo {pages} {1730} (\bibinfo {year} {2019})}\BibitemShut {NoStop}%
\bibitem [{\citenamefont {Dogra}\ \emph {et~al.}(2019)\citenamefont {Dogra},
  \citenamefont {Landini}, \citenamefont {Kroeger}, \citenamefont {Hruby},
  \citenamefont {Donner},\ and\ \citenamefont {Esslinger}}]{Nishant2019}%
  \BibitemOpen
  \bibfield  {author} {\bibinfo {author} {\bibfnamefont {N.}~\bibnamefont
  {Dogra}}, \bibinfo {author} {\bibfnamefont {M.}~\bibnamefont {Landini}},
  \bibinfo {author} {\bibfnamefont {K.}~\bibnamefont {Kroeger}}, \bibinfo
  {author} {\bibfnamefont {L.}~\bibnamefont {Hruby}}, \bibinfo {author}
  {\bibfnamefont {T.}~\bibnamefont {Donner}},\ and\ \bibinfo {author}
  {\bibfnamefont {T.}~\bibnamefont {Esslinger}},\ }\bibfield  {title} {\bibinfo
  {title} {Dissipation-induced structural instability and chiral dynamics in a
  quantum gas},\ }\href {https://doi.org/10.1126/science.aaw4465} {\bibfield
  {journal} {\bibinfo  {journal} {Science}\ }\textbf {\bibinfo {volume}
  {366}},\ \bibinfo {pages} {1496} (\bibinfo {year} {2019})}\BibitemShut
  {NoStop}%
\bibitem [{\citenamefont {Bu\ifmmode~\check{c}\else
  \v{c}\fi{}a}(2023)}]{Buca2023prx}%
  \BibitemOpen
  \bibfield  {author} {\bibinfo {author} {\bibfnamefont {B.}~\bibnamefont
  {Bu\ifmmode~\check{c}\else \v{c}\fi{}a}},\ }\bibfield  {title} {\bibinfo
  {title} {Unified theory of local quantum many-body dynamics: Eigenoperator
  thermalization theorems},\ }\href
  {https://doi.org/10.1103/PhysRevX.13.031013} {\bibfield  {journal} {\bibinfo
  {journal} {Phys. Rev. X}\ }\textbf {\bibinfo {volume} {13}},\ \bibinfo
  {pages} {031013} (\bibinfo {year} {2023})}\BibitemShut {NoStop}%
\bibitem [{\citenamefont {Greenberger}\ \emph {et~al.}(1990)\citenamefont
  {Greenberger}, \citenamefont {Horne}, \citenamefont {Shimony},\ and\
  \citenamefont {Zeilinger}}]{GHZ1990}%
  \BibitemOpen
  \bibfield  {author} {\bibinfo {author} {\bibfnamefont {D.~M.}\ \bibnamefont
  {Greenberger}}, \bibinfo {author} {\bibfnamefont {M.~A.}\ \bibnamefont
  {Horne}}, \bibinfo {author} {\bibfnamefont {A.}~\bibnamefont {Shimony}},\
  and\ \bibinfo {author} {\bibfnamefont {A.}~\bibnamefont {Zeilinger}},\
  }\bibfield  {title} {\bibinfo {title} {Bell’s theorem without
  inequalities},\ }\href {https://doi.org/10.1119/1.16243} {\bibfield
  {journal} {\bibinfo  {journal} {American Journal of Physics}\ }\textbf
  {\bibinfo {volume} {58}},\ \bibinfo {pages} {1131} (\bibinfo {year}
  {1990})}\BibitemShut {NoStop}%
\bibitem [{\citenamefont {Horodecki}\ \emph {et~al.}(2009)\citenamefont
  {Horodecki}, \citenamefont {Horodecki}, \citenamefont {Horodecki},\ and\
  \citenamefont {Horodecki}}]{Horodecki2009RMP}%
  \BibitemOpen
  \bibfield  {author} {\bibinfo {author} {\bibfnamefont {R.}~\bibnamefont
  {Horodecki}}, \bibinfo {author} {\bibfnamefont {P.}~\bibnamefont
  {Horodecki}}, \bibinfo {author} {\bibfnamefont {M.}~\bibnamefont
  {Horodecki}},\ and\ \bibinfo {author} {\bibfnamefont {K.}~\bibnamefont
  {Horodecki}},\ }\bibfield  {title} {\bibinfo {title} {Quantum entanglement},\
  }\href {https://doi.org/10.1103/RevModPhys.81.865} {\bibfield  {journal}
  {\bibinfo  {journal} {Rev. Mod. Phys.}\ }\textbf {\bibinfo {volume} {81}},\
  \bibinfo {pages} {865} (\bibinfo {year} {2009})}\BibitemShut {NoStop}%
\bibitem [{\citenamefont {Ben-Ami}\ \emph {et~al.}(2025)\citenamefont
  {Ben-Ami}, \citenamefont {Heyl},\ and\ \citenamefont
  {Moessner}}]{Benami2025}%
  \BibitemOpen
  \bibfield  {author} {\bibinfo {author} {\bibfnamefont {T.}~\bibnamefont
  {Ben-Ami}}, \bibinfo {author} {\bibfnamefont {M.}~\bibnamefont {Heyl}},\ and\
  \bibinfo {author} {\bibfnamefont {R.}~\bibnamefont {Moessner}},\ }\bibfield
  {title} {\bibinfo {title} {Many-body cages: disorder-free glassiness from
  flat bands in fock space, and many-body rabi oscillations},\ }\href
  {https://arxiv.org/abs/2504.13086} {\bibfield  {journal} {\bibinfo  {journal}
  {arXiv preprint arXiv:2504.13086}\ } (\bibinfo {year} {2025})}\BibitemShut
  {NoStop}%
\bibitem [{\citenamefont {Reed}\ \emph {et~al.}(2012)\citenamefont {Reed},
  \citenamefont {DiCarlo}, \citenamefont {Nigg}, \citenamefont {Sun},
  \citenamefont {Frunzio}, \citenamefont {Girvin},\ and\ \citenamefont
  {Schoelkopf}}]{Reed2012}%
  \BibitemOpen
  \bibfield  {author} {\bibinfo {author} {\bibfnamefont {M.~D.}\ \bibnamefont
  {Reed}}, \bibinfo {author} {\bibfnamefont {L.}~\bibnamefont {DiCarlo}},
  \bibinfo {author} {\bibfnamefont {S.~E.}\ \bibnamefont {Nigg}}, \bibinfo
  {author} {\bibfnamefont {L.}~\bibnamefont {Sun}}, \bibinfo {author}
  {\bibfnamefont {L.}~\bibnamefont {Frunzio}}, \bibinfo {author} {\bibfnamefont
  {S.~M.}\ \bibnamefont {Girvin}},\ and\ \bibinfo {author} {\bibfnamefont
  {R.~J.}\ \bibnamefont {Schoelkopf}},\ }\bibfield  {title} {\bibinfo {title}
  {Realization of three-qubit quantum error correction with superconducting
  circuits},\ }\href {https://doi.org/10.1038/nature10786} {\bibfield
  {journal} {\bibinfo  {journal} {Nature}\ }\textbf {\bibinfo {volume} {482}},\
  \bibinfo {pages} {382} (\bibinfo {year} {2012})}\BibitemShut {NoStop}%
\bibitem [{\citenamefont {Tóth}\ and\ \citenamefont
  {Apellaniz}(2014)}]{Toth2014}%
  \BibitemOpen
  \bibfield  {author} {\bibinfo {author} {\bibfnamefont {G.}~\bibnamefont
  {Tóth}}\ and\ \bibinfo {author} {\bibfnamefont {I.}~\bibnamefont
  {Apellaniz}},\ }\bibfield  {title} {\bibinfo {title} {Quantum metrology from
  a quantum information science perspective},\ }\href
  {https://doi.org/10.1088/1751-8113/47/42/424006} {\bibfield  {journal}
  {\bibinfo  {journal} {Journal of Physics A: Mathematical and Theoretical}\
  }\textbf {\bibinfo {volume} {47}},\ \bibinfo {pages} {424006} (\bibinfo
  {year} {2014})}\BibitemShut {NoStop}%
\bibitem [{\citenamefont {Bouwmeester}\ \emph {et~al.}(1999)\citenamefont
  {Bouwmeester}, \citenamefont {Pan}, \citenamefont {Daniell}, \citenamefont
  {Weinfurter},\ and\ \citenamefont {Zeilinger}}]{Pan1999}%
  \BibitemOpen
  \bibfield  {author} {\bibinfo {author} {\bibfnamefont {D.}~\bibnamefont
  {Bouwmeester}}, \bibinfo {author} {\bibfnamefont {J.-W.}\ \bibnamefont
  {Pan}}, \bibinfo {author} {\bibfnamefont {M.}~\bibnamefont {Daniell}},
  \bibinfo {author} {\bibfnamefont {H.}~\bibnamefont {Weinfurter}},\ and\
  \bibinfo {author} {\bibfnamefont {A.}~\bibnamefont {Zeilinger}},\ }\bibfield
  {title} {\bibinfo {title} {Observation of three-photon
  greenberger-horne-zeilinger entanglement},\ }\href
  {https://doi.org/10.1103/PhysRevLett.82.1345} {\bibfield  {journal} {\bibinfo
   {journal} {Phys. Rev. Lett.}\ }\textbf {\bibinfo {volume} {82}},\ \bibinfo
  {pages} {1345} (\bibinfo {year} {1999})}\BibitemShut {NoStop}%
\bibitem [{\citenamefont {DiCarlo}\ \emph {et~al.}(2010)\citenamefont
  {DiCarlo}, \citenamefont {Reed}, \citenamefont {Sun}, \citenamefont
  {Johnson}, \citenamefont {Chow}, \citenamefont {Gambetta}, \citenamefont
  {Frunzio}, \citenamefont {Girvin}, \citenamefont {Devoret},\ and\
  \citenamefont {Schoelkopf}}]{DiCarlo2010}%
  \BibitemOpen
  \bibfield  {author} {\bibinfo {author} {\bibfnamefont {L.}~\bibnamefont
  {DiCarlo}}, \bibinfo {author} {\bibfnamefont {M.~D.}\ \bibnamefont {Reed}},
  \bibinfo {author} {\bibfnamefont {L.}~\bibnamefont {Sun}}, \bibinfo {author}
  {\bibfnamefont {B.~R.}\ \bibnamefont {Johnson}}, \bibinfo {author}
  {\bibfnamefont {J.~M.}\ \bibnamefont {Chow}}, \bibinfo {author}
  {\bibfnamefont {J.~M.}\ \bibnamefont {Gambetta}}, \bibinfo {author}
  {\bibfnamefont {L.}~\bibnamefont {Frunzio}}, \bibinfo {author} {\bibfnamefont
  {S.~M.}\ \bibnamefont {Girvin}}, \bibinfo {author} {\bibfnamefont {M.~H.}\
  \bibnamefont {Devoret}},\ and\ \bibinfo {author} {\bibfnamefont {R.~J.}\
  \bibnamefont {Schoelkopf}},\ }\bibfield  {title} {\bibinfo {title}
  {Preparation and measurement of three-qubit entanglement in a superconducting
  circuit},\ }\href {https://doi.org/10.1038/nature09416} {\bibfield  {journal}
  {\bibinfo  {journal} {Nature}\ }\textbf {\bibinfo {volume} {467}},\ \bibinfo
  {pages} {574} (\bibinfo {year} {2010})}\BibitemShut {NoStop}%
\bibitem [{\citenamefont {Monz}\ \emph {et~al.}(2011)\citenamefont {Monz},
  \citenamefont {Schindler}, \citenamefont {Barreiro}, \citenamefont {Chwalla},
  \citenamefont {Nigg}, \citenamefont {Coish}, \citenamefont {Harlander},
  \citenamefont {H\"ansel}, \citenamefont {Hennrich},\ and\ \citenamefont
  {Blatt}}]{Monz2011}%
  \BibitemOpen
  \bibfield  {author} {\bibinfo {author} {\bibfnamefont {T.}~\bibnamefont
  {Monz}}, \bibinfo {author} {\bibfnamefont {P.}~\bibnamefont {Schindler}},
  \bibinfo {author} {\bibfnamefont {J.~T.}\ \bibnamefont {Barreiro}}, \bibinfo
  {author} {\bibfnamefont {M.}~\bibnamefont {Chwalla}}, \bibinfo {author}
  {\bibfnamefont {D.}~\bibnamefont {Nigg}}, \bibinfo {author} {\bibfnamefont
  {W.~A.}\ \bibnamefont {Coish}}, \bibinfo {author} {\bibfnamefont
  {M.}~\bibnamefont {Harlander}}, \bibinfo {author} {\bibfnamefont
  {W.}~\bibnamefont {H\"ansel}}, \bibinfo {author} {\bibfnamefont
  {M.}~\bibnamefont {Hennrich}},\ and\ \bibinfo {author} {\bibfnamefont
  {R.}~\bibnamefont {Blatt}},\ }\bibfield  {title} {\bibinfo {title} {14-qubit
  entanglement: Creation and coherence},\ }\href
  {https://doi.org/10.1103/PhysRevLett.106.130506} {\bibfield  {journal}
  {\bibinfo  {journal} {Phys. Rev. Lett.}\ }\textbf {\bibinfo {volume} {106}},\
  \bibinfo {pages} {130506} (\bibinfo {year} {2011})}\BibitemShut {NoStop}%
\bibitem [{\citenamefont {Zhu}\ \emph {et~al.}(2023)\citenamefont {Zhu},
  \citenamefont {Tantivasadakarn}, \citenamefont {Vishwanath}, \citenamefont
  {Trebst},\ and\ \citenamefont {Verresen}}]{Nishimori}%
  \BibitemOpen
  \bibfield  {author} {\bibinfo {author} {\bibfnamefont {G.-Y.}\ \bibnamefont
  {Zhu}}, \bibinfo {author} {\bibfnamefont {N.}~\bibnamefont
  {Tantivasadakarn}}, \bibinfo {author} {\bibfnamefont {A.}~\bibnamefont
  {Vishwanath}}, \bibinfo {author} {\bibfnamefont {S.}~\bibnamefont {Trebst}},\
  and\ \bibinfo {author} {\bibfnamefont {R.}~\bibnamefont {Verresen}},\
  }\bibfield  {title} {\bibinfo {title} {Nishimori's cat: Stable long-range
  entanglement from finite-depth unitaries and weak measurements},\ }\href
  {https://doi.org/10.1103/PhysRevLett.131.200201} {\bibfield  {journal}
  {\bibinfo  {journal} {Phys. Rev. Lett.}\ }\textbf {\bibinfo {volume} {131}},\
  \bibinfo {pages} {200201} (\bibinfo {year} {2023})}\BibitemShut {NoStop}%
\bibitem [{\citenamefont {D'Alessio}\ \emph {et~al.}(2016)\citenamefont
  {D'Alessio}, \citenamefont {Kafri}, \citenamefont {Polkovnikov},\ and\
  \citenamefont {and}}]{d2016quantum}%
  \BibitemOpen
  \bibfield  {author} {\bibinfo {author} {\bibfnamefont {L.}~\bibnamefont
  {D'Alessio}}, \bibinfo {author} {\bibfnamefont {Y.}~\bibnamefont {Kafri}},
  \bibinfo {author} {\bibfnamefont {A.}~\bibnamefont {Polkovnikov}},\ and\
  \bibinfo {author} {\bibfnamefont {M.~R.}\ \bibnamefont {and}},\ }\bibfield
  {title} {\bibinfo {title} {From quantum chaos and eigenstate thermalization
  to statistical mechanics and thermodynamics},\ }\href
  {https://doi.org/10.1080/00018732.2016.1198134} {\bibfield  {journal}
  {\bibinfo  {journal} {Advances in Physics}\ }\textbf {\bibinfo {volume}
  {65}},\ \bibinfo {pages} {239} (\bibinfo {year} {2016})}\BibitemShut
  {NoStop}%
\bibitem [{\citenamefont {Krovi}\ and\ \citenamefont {Brun}(2006)}]{Krovi2006}%
  \BibitemOpen
  \bibfield  {author} {\bibinfo {author} {\bibfnamefont {H.}~\bibnamefont
  {Krovi}}\ and\ \bibinfo {author} {\bibfnamefont {T.~A.}\ \bibnamefont
  {Brun}},\ }\bibfield  {title} {\bibinfo {title} {Quantum walks with infinite
  hitting times},\ }\href {https://doi.org/10.1103/PhysRevA.74.042334}
  {\bibfield  {journal} {\bibinfo  {journal} {Phys. Rev. A}\ }\textbf {\bibinfo
  {volume} {74}},\ \bibinfo {pages} {042334} (\bibinfo {year}
  {2006})}\BibitemShut {NoStop}%
\bibitem [{\citenamefont {Gr{\"{u}}nbaum}\ \emph {et~al.}(2013)\citenamefont
  {Gr{\"{u}}nbaum}, \citenamefont {Vel{\'{a}}zquez}, \citenamefont {Werner},\
  and\ \citenamefont {Werner}}]{Gruenbaum2013}%
  \BibitemOpen
  \bibfield  {author} {\bibinfo {author} {\bibfnamefont {F.~A.}\ \bibnamefont
  {Gr{\"{u}}nbaum}}, \bibinfo {author} {\bibfnamefont {L.}~\bibnamefont
  {Vel{\'{a}}zquez}}, \bibinfo {author} {\bibfnamefont {A.~H.}\ \bibnamefont
  {Werner}},\ and\ \bibinfo {author} {\bibfnamefont {R.~F.}\ \bibnamefont
  {Werner}},\ }\bibfield  {title} {\bibinfo {title} {Recurrence for discrete
  time unitary evolutions},\ }\href {https://doi.org/10.1007/s00220-012-1645-2}
  {\bibfield  {journal} {\bibinfo  {journal} {Communications in Mathematical
  Physics}\ }\textbf {\bibinfo {volume} {320}},\ \bibinfo {pages} {543}
  (\bibinfo {year} {2013})}\BibitemShut {NoStop}%
\bibitem [{\citenamefont {Dhar}\ \emph {et~al.}(2015)\citenamefont {Dhar},
  \citenamefont {Dasgupta}, \citenamefont {Dhar},\ and\ \citenamefont
  {Sen}}]{Dhar2015}%
  \BibitemOpen
  \bibfield  {author} {\bibinfo {author} {\bibfnamefont {S.}~\bibnamefont
  {Dhar}}, \bibinfo {author} {\bibfnamefont {S.}~\bibnamefont {Dasgupta}},
  \bibinfo {author} {\bibfnamefont {A.}~\bibnamefont {Dhar}},\ and\ \bibinfo
  {author} {\bibfnamefont {D.}~\bibnamefont {Sen}},\ }\bibfield  {title}
  {\bibinfo {title} {Detection of a quantum particle on a lattice under
  repeated projective measurements},\ }\href
  {https://doi.org/10.1103/PhysRevA.91.062115} {\bibfield  {journal} {\bibinfo
  {journal} {Phys. Rev. A}\ }\textbf {\bibinfo {volume} {91}},\ \bibinfo
  {pages} {062115} (\bibinfo {year} {2015})}\BibitemShut {NoStop}%
\bibitem [{\citenamefont {Friedman}\ \emph {et~al.}(2017)\citenamefont
  {Friedman}, \citenamefont {Kessler},\ and\ \citenamefont
  {Barkai}}]{Friedman2017a}%
  \BibitemOpen
  \bibfield  {author} {\bibinfo {author} {\bibfnamefont {H.}~\bibnamefont
  {Friedman}}, \bibinfo {author} {\bibfnamefont {D.~A.}\ \bibnamefont
  {Kessler}},\ and\ \bibinfo {author} {\bibfnamefont {E.}~\bibnamefont
  {Barkai}},\ }\bibfield  {title} {\bibinfo {title} {Quantum walks: The first
  detected passage time problem},\ }\href
  {https://doi.org/10.1103/PhysRevE.95.032141} {\bibfield  {journal} {\bibinfo
  {journal} {Phys. Rev. E}\ }\textbf {\bibinfo {volume} {95}},\ \bibinfo
  {pages} {032141} (\bibinfo {year} {2017})}\BibitemShut {NoStop}%
\bibitem [{\citenamefont {Lahiri}\ and\ \citenamefont
  {Dhar}(2019)}]{Lahiri2019}%
  \BibitemOpen
  \bibfield  {author} {\bibinfo {author} {\bibfnamefont {S.}~\bibnamefont
  {Lahiri}}\ and\ \bibinfo {author} {\bibfnamefont {A.}~\bibnamefont {Dhar}},\
  }\bibfield  {title} {\bibinfo {title} {Return to the origin problem for a
  particle on a one-dimensional lattice with quasi-zeno dynamics},\ }\href
  {https://doi.org/10.1103/PhysRevA.99.012101} {\bibfield  {journal} {\bibinfo
  {journal} {Phys. Rev. A}\ }\textbf {\bibinfo {volume} {99}},\ \bibinfo
  {pages} {012101} (\bibinfo {year} {2019})}\BibitemShut {NoStop}%
\bibitem [{\citenamefont {Dubey}\ \emph {et~al.}(2021)\citenamefont {Dubey},
  \citenamefont {Bernardin},\ and\ \citenamefont {Dhar}}]{dubey2021quantum}%
  \BibitemOpen
  \bibfield  {author} {\bibinfo {author} {\bibfnamefont {V.}~\bibnamefont
  {Dubey}}, \bibinfo {author} {\bibfnamefont {C.}~\bibnamefont {Bernardin}},\
  and\ \bibinfo {author} {\bibfnamefont {A.}~\bibnamefont {Dhar}},\ }\bibfield
  {title} {\bibinfo {title} {Quantum dynamics under continuous projective
  measurements: Non-hermitian description and the continuum-space limit},\
  }\href {https://journals.aps.org/pra/abstract/10.1103/PhysRevA.103.032221}
  {\bibfield  {journal} {\bibinfo  {journal} {Phys. Rev. A}\ }\textbf {\bibinfo
  {volume} {103}},\ \bibinfo {pages} {032221} (\bibinfo {year}
  {2021})}\BibitemShut {NoStop}%
\bibitem [{\citenamefont {Tornow}\ and\ \citenamefont
  {Ziegler}(2023)}]{Sabine2022}%
  \BibitemOpen
  \bibfield  {author} {\bibinfo {author} {\bibfnamefont {S.}~\bibnamefont
  {Tornow}}\ and\ \bibinfo {author} {\bibfnamefont {K.}~\bibnamefont
  {Ziegler}},\ }\bibfield  {title} {\bibinfo {title} {Measurement-induced
  quantum walks on an ibm quantum computer},\ }\href
  {https://doi.org/10.1103/PhysRevResearch.5.033089} {\bibfield  {journal}
  {\bibinfo  {journal} {Phys. Rev. Res.}\ }\textbf {\bibinfo {volume} {5}},\
  \bibinfo {pages} {033089} (\bibinfo {year} {2023})}\BibitemShut {NoStop}%
\bibitem [{\citenamefont {Yin}\ and\ \citenamefont {Barkai}(2023)}]{Ruoyu2023}%
  \BibitemOpen
  \bibfield  {author} {\bibinfo {author} {\bibfnamefont {R.}~\bibnamefont
  {Yin}}\ and\ \bibinfo {author} {\bibfnamefont {E.}~\bibnamefont {Barkai}},\
  }\bibfield  {title} {\bibinfo {title} {Restart expedites quantum walk hitting
  times},\ }\href {https://doi.org/10.1103/PhysRevLett.130.050802} {\bibfield
  {journal} {\bibinfo  {journal} {Phys. Rev. Lett.}\ }\textbf {\bibinfo
  {volume} {130}},\ \bibinfo {pages} {050802} (\bibinfo {year}
  {2023})}\BibitemShut {NoStop}%
\bibitem [{\citenamefont {Wang}\ \emph {et~al.}(2023)\citenamefont {Wang},
  \citenamefont {Yin}, \citenamefont {Dou}, \citenamefont {Zhang},\ and\
  \citenamefont {Song}}]{Yajing2023}%
  \BibitemOpen
  \bibfield  {author} {\bibinfo {author} {\bibfnamefont {Y.-J.}\ \bibnamefont
  {Wang}}, \bibinfo {author} {\bibfnamefont {R.-Y.}\ \bibnamefont {Yin}},
  \bibinfo {author} {\bibfnamefont {L.-Y.}\ \bibnamefont {Dou}}, \bibinfo
  {author} {\bibfnamefont {A.-N.}\ \bibnamefont {Zhang}},\ and\ \bibinfo
  {author} {\bibfnamefont {X.-B.}\ \bibnamefont {Song}},\ }\bibfield  {title}
  {\bibinfo {title} {Quantum first detection of a quantum walker on a perturbed
  ring},\ }\href {https://doi.org/10.1103/PhysRevResearch.5.013202} {\bibfield
  {journal} {\bibinfo  {journal} {Phys. Rev. Res.}\ }\textbf {\bibinfo {volume}
  {5}},\ \bibinfo {pages} {013202} (\bibinfo {year} {2023})}\BibitemShut
  {NoStop}%
\bibitem [{\citenamefont {Ni}\ and\ \citenamefont {Zheng}(2023)}]{Zhenbo2023}%
  \BibitemOpen
  \bibfield  {author} {\bibinfo {author} {\bibfnamefont {Z.}~\bibnamefont
  {Ni}}\ and\ \bibinfo {author} {\bibfnamefont {Y.}~\bibnamefont {Zheng}},\
  }\bibfield  {title} {\bibinfo {title} {First detection and tunneling time of
  a quantum walk},\ }\href {https://doi.org/10.3390/e25081231} {\bibfield
  {journal} {\bibinfo  {journal} {Entropy}\ }\textbf {\bibinfo {volume} {25}},\
  \bibinfo {pages} {1231} (\bibinfo {year} {2023})}\BibitemShut {NoStop}%
\bibitem [{\citenamefont {Yin}\ \emph {et~al.}(2025)\citenamefont {Yin},
  \citenamefont {Wang}, \citenamefont {Tornow},\ and\ \citenamefont
  {Barkai}}]{yin2024restart}%
  \BibitemOpen
  \bibfield  {author} {\bibinfo {author} {\bibfnamefont {R.}~\bibnamefont
  {Yin}}, \bibinfo {author} {\bibfnamefont {Q.}~\bibnamefont {Wang}}, \bibinfo
  {author} {\bibfnamefont {S.}~\bibnamefont {Tornow}},\ and\ \bibinfo {author}
  {\bibfnamefont {E.}~\bibnamefont {Barkai}},\ }\bibfield  {title} {\bibinfo
  {title} {Restart uncertainty relation for monitored quantum dynamics},\
  }\href {https://doi.org/10.1073/pnas.2402912121} {\bibfield  {journal}
  {\bibinfo  {journal} {Proceedings of the National Academy of Sciences}\
  }\textbf {\bibinfo {volume} {122}},\ \bibinfo {pages} {e2402912121} (\bibinfo
  {year} {2025})}\BibitemShut {NoStop}%
\bibitem [{\citenamefont {Roy}\ \emph {et~al.}(2025)\citenamefont {Roy},
  \citenamefont {Gupta},\ and\ \citenamefont {Morigi}}]{Giovanna2025causality}%
  \BibitemOpen
  \bibfield  {author} {\bibinfo {author} {\bibfnamefont {S.}~\bibnamefont
  {Roy}}, \bibinfo {author} {\bibfnamefont {S.}~\bibnamefont {Gupta}},\ and\
  \bibinfo {author} {\bibfnamefont {G.}~\bibnamefont {Morigi}},\ }\bibfield
  {title} {\bibinfo {title} {Causality, localisation, and universality of
  monitored quantum walks with long-range hopping},\ }\href
  {https://arxiv.org/abs/2504.12053} {\bibfield  {journal} {\bibinfo  {journal}
  {arXiv preprint arXiv:2504.12053}\ } (\bibinfo {year} {2025})}\BibitemShut
  {NoStop}%
\bibitem [{\citenamefont {Schecter}\ and\ \citenamefont
  {Iadecola}(2019)}]{Thomas2019}%
  \BibitemOpen
  \bibfield  {author} {\bibinfo {author} {\bibfnamefont {M.}~\bibnamefont
  {Schecter}}\ and\ \bibinfo {author} {\bibfnamefont {T.}~\bibnamefont
  {Iadecola}},\ }\bibfield  {title} {\bibinfo {title} {Weak ergodicity breaking
  and quantum many-body scars in spin-1 $xy$ magnets},\ }\href
  {https://doi.org/10.1103/PhysRevLett.123.147201} {\bibfield  {journal}
  {\bibinfo  {journal} {Phys. Rev. Lett.}\ }\textbf {\bibinfo {volume} {123}},\
  \bibinfo {pages} {147201} (\bibinfo {year} {2019})}\BibitemShut {NoStop}%
\bibitem [{\citenamefont {Moudgalya}\ \emph {et~al.}(2018)\citenamefont
  {Moudgalya}, \citenamefont {Rachel}, \citenamefont {Bernevig},\ and\
  \citenamefont {Regnault}}]{Bernevig2018a}%
  \BibitemOpen
  \bibfield  {author} {\bibinfo {author} {\bibfnamefont {S.}~\bibnamefont
  {Moudgalya}}, \bibinfo {author} {\bibfnamefont {S.}~\bibnamefont {Rachel}},
  \bibinfo {author} {\bibfnamefont {B.~A.}\ \bibnamefont {Bernevig}},\ and\
  \bibinfo {author} {\bibfnamefont {N.}~\bibnamefont {Regnault}},\ }\bibfield
  {title} {\bibinfo {title} {Exact excited states of nonintegrable models},\
  }\href {https://doi.org/10.1103/PhysRevB.98.235155} {\bibfield  {journal}
  {\bibinfo  {journal} {Phys. Rev. B}\ }\textbf {\bibinfo {volume} {98}},\
  \bibinfo {pages} {235155} (\bibinfo {year} {2018})}\BibitemShut {NoStop}%
\bibitem [{\citenamefont {Moudgalya}\ \emph {et~al.}(2020)\citenamefont
  {Moudgalya}, \citenamefont {Regnault},\ and\ \citenamefont
  {Bernevig}}]{Sanjay2020}%
  \BibitemOpen
  \bibfield  {author} {\bibinfo {author} {\bibfnamefont {S.}~\bibnamefont
  {Moudgalya}}, \bibinfo {author} {\bibfnamefont {N.}~\bibnamefont
  {Regnault}},\ and\ \bibinfo {author} {\bibfnamefont {B.~A.}\ \bibnamefont
  {Bernevig}},\ }\bibfield  {title} {\bibinfo {title}
  {$\ensuremath{\eta}$-pairing in hubbard models: From spectrum generating
  algebras to quantum many-body scars},\ }\href
  {https://doi.org/10.1103/PhysRevB.102.085140} {\bibfield  {journal} {\bibinfo
   {journal} {Phys. Rev. B}\ }\textbf {\bibinfo {volume} {102}},\ \bibinfo
  {pages} {085140} (\bibinfo {year} {2020})}\BibitemShut {NoStop}%
\bibitem [{\citenamefont {Mark}\ \emph {et~al.}(2020)\citenamefont {Mark},
  \citenamefont {Lin},\ and\ \citenamefont {Motrunich}}]{Mark2020a}%
  \BibitemOpen
  \bibfield  {author} {\bibinfo {author} {\bibfnamefont {D.~K.}\ \bibnamefont
  {Mark}}, \bibinfo {author} {\bibfnamefont {C.-J.}\ \bibnamefont {Lin}},\ and\
  \bibinfo {author} {\bibfnamefont {O.~I.}\ \bibnamefont {Motrunich}},\
  }\bibfield  {title} {\bibinfo {title} {Unified structure for exact towers of
  scar states in the affleck-kennedy-lieb-tasaki and other models},\ }\href
  {https://doi.org/10.1103/PhysRevB.101.195131} {\bibfield  {journal} {\bibinfo
   {journal} {Phys. Rev. B}\ }\textbf {\bibinfo {volume} {101}},\ \bibinfo
  {pages} {195131} (\bibinfo {year} {2020})}\BibitemShut {NoStop}%
\bibitem [{\citenamefont {Chandran}\ \emph {et~al.}(2023)\citenamefont
  {Chandran}, \citenamefont {Iadecola}, \citenamefont {Khemani},\ and\
  \citenamefont {Moessner}}]{Roderich2023Review}%
  \BibitemOpen
  \bibfield  {author} {\bibinfo {author} {\bibfnamefont {A.}~\bibnamefont
  {Chandran}}, \bibinfo {author} {\bibfnamefont {T.}~\bibnamefont {Iadecola}},
  \bibinfo {author} {\bibfnamefont {V.}~\bibnamefont {Khemani}},\ and\ \bibinfo
  {author} {\bibfnamefont {R.}~\bibnamefont {Moessner}},\ }\bibfield  {title}
  {\bibinfo {title} {Quantum many-body scars: A quasiparticle perspective},\
  }\href
  {https://doi.org/https://doi.org/10.1146/annurev-conmatphys-031620-101617}
  {\bibfield  {journal} {\bibinfo  {journal} {Annual Review of Condensed Matter
  Physics}\ }\textbf {\bibinfo {volume} {14}},\ \bibinfo {pages} {443}
  (\bibinfo {year} {2023})}\BibitemShut {NoStop}%
\bibitem [{\citenamefont {Walter}\ \emph {et~al.}(2025)\citenamefont {Walter},
  \citenamefont {Perfetto},\ and\ \citenamefont
  {Gambassi}}]{walter2025thermodynamic}%
  \BibitemOpen
  \bibfield  {author} {\bibinfo {author} {\bibfnamefont {B.}~\bibnamefont
  {Walter}}, \bibinfo {author} {\bibfnamefont {G.}~\bibnamefont {Perfetto}},\
  and\ \bibinfo {author} {\bibfnamefont {A.}~\bibnamefont {Gambassi}},\
  }\bibfield  {title} {\bibinfo {title} {Thermodynamic phases in first detected
  return times of quantum many-body systems},\ }\href
  {https://doi.org/10.1103/PhysRevA.111.L040202} {\bibfield  {journal}
  {\bibinfo  {journal} {Phys. Rev. A}\ }\textbf {\bibinfo {volume} {111}},\
  \bibinfo {pages} {L040202} (\bibinfo {year} {2025})}\BibitemShut {NoStop}%
\bibitem [{\citenamefont {AI}\ and\ \citenamefont
  {Collaborators}(2023)}]{GoogleQuantumAI2023}%
  \BibitemOpen
  \bibfield  {author} {\bibinfo {author} {\bibfnamefont {G.~Q.}\ \bibnamefont
  {AI}}\ and\ \bibinfo {author} {\bibnamefont {Collaborators}},\ }\bibfield
  {title} {\bibinfo {title} {Measurement-induced entanglement and teleportation
  on a noisy quantum processor},\ }\href
  {https://doi.org/10.1038/s41586-023-06505-7} {\bibfield  {journal} {\bibinfo
  {journal} {Nature}\ }\textbf {\bibinfo {volume} {622}},\ \bibinfo {pages}
  {481} (\bibinfo {year} {2023})}\BibitemShut {NoStop}%
\bibitem [{\citenamefont {Koh}\ \emph {et~al.}(2023)\citenamefont {Koh},
  \citenamefont {Sun}, \citenamefont {Motta},\ and\ \citenamefont
  {Minnich}}]{koh2023measurement}%
  \BibitemOpen
  \bibfield  {author} {\bibinfo {author} {\bibfnamefont {J.~M.}\ \bibnamefont
  {Koh}}, \bibinfo {author} {\bibfnamefont {S.-N.}\ \bibnamefont {Sun}},
  \bibinfo {author} {\bibfnamefont {M.}~\bibnamefont {Motta}},\ and\ \bibinfo
  {author} {\bibfnamefont {A.~J.}\ \bibnamefont {Minnich}},\ }\bibfield
  {title} {\bibinfo {title} {Measurement-induced entanglement phase transition
  on a superconducting quantum processor with mid-circuit readout},\ }\href
  {https://doi.org/10.1038/s41567-023-02076-6} {\bibfield  {journal} {\bibinfo
  {journal} {Nature Physics}\ }\textbf {\bibinfo {volume} {19}},\ \bibinfo
  {pages} {1314} (\bibinfo {year} {2023})}\BibitemShut {NoStop}%
\bibitem [{\citenamefont {Thiel}\ \emph {et~al.}(2020)\citenamefont {Thiel},
  \citenamefont {Mualem}, \citenamefont {Meidan}, \citenamefont {Barkai},\ and\
  \citenamefont {Kessler}}]{Thiel2020D}%
  \BibitemOpen
  \bibfield  {author} {\bibinfo {author} {\bibfnamefont {F.}~\bibnamefont
  {Thiel}}, \bibinfo {author} {\bibfnamefont {I.}~\bibnamefont {Mualem}},
  \bibinfo {author} {\bibfnamefont {D.}~\bibnamefont {Meidan}}, \bibinfo
  {author} {\bibfnamefont {E.}~\bibnamefont {Barkai}},\ and\ \bibinfo {author}
  {\bibfnamefont {D.~A.}\ \bibnamefont {Kessler}},\ }\bibfield  {title}
  {\bibinfo {title} {Dark states of quantum search cause imperfect detection},\
  }\href {https://doi.org/10.1103/PhysRevResearch.2.043107} {\bibfield
  {journal} {\bibinfo  {journal} {Phys. Rev. Research}\ }\textbf {\bibinfo
  {volume} {2}},\ \bibinfo {pages} {043107} (\bibinfo {year}
  {2020})}\BibitemShut {NoStop}%
\bibitem [{\citenamefont {Liu}\ \emph {et~al.}(2022)\citenamefont {Liu},
  \citenamefont {Ziegler}, \citenamefont {Kessler},\ and\ \citenamefont
  {Barkai}}]{Liu2022a}%
  \BibitemOpen
  \bibfield  {author} {\bibinfo {author} {\bibfnamefont {Q.}~\bibnamefont
  {Liu}}, \bibinfo {author} {\bibfnamefont {K.}~\bibnamefont {Ziegler}},
  \bibinfo {author} {\bibfnamefont {D.~A.}\ \bibnamefont {Kessler}},\ and\
  \bibinfo {author} {\bibfnamefont {E.}~\bibnamefont {Barkai}},\ }\bibfield
  {title} {\bibinfo {title} {Driving quantum systems with periodic conditional
  measurements},\ }\href {https://doi.org/10.1103/PhysRevResearch.4.023129}
  {\bibfield  {journal} {\bibinfo  {journal} {Phys. Rev. Res.}\ }\textbf
  {\bibinfo {volume} {4}},\ \bibinfo {pages} {023129} (\bibinfo {year}
  {2022})}\BibitemShut {NoStop}%
\bibitem [{\citenamefont {Yin}\ \emph {et~al.}(2019)\citenamefont {Yin},
  \citenamefont {Ziegler}, \citenamefont {Thiel},\ and\ \citenamefont
  {Barkai}}]{yin2019}%
  \BibitemOpen
  \bibfield  {author} {\bibinfo {author} {\bibfnamefont {R.}~\bibnamefont
  {Yin}}, \bibinfo {author} {\bibfnamefont {K.}~\bibnamefont {Ziegler}},
  \bibinfo {author} {\bibfnamefont {F.}~\bibnamefont {Thiel}},\ and\ \bibinfo
  {author} {\bibfnamefont {E.}~\bibnamefont {Barkai}},\ }\bibfield  {title}
  {\bibinfo {title} {Large fluctuations of the first detected quantum return
  time},\ }\href {https://doi.org/10.1103/PhysRevResearch.1.033086} {\bibfield
  {journal} {\bibinfo  {journal} {Phys. Rev. Research}\ }\textbf {\bibinfo
  {volume} {1}},\ \bibinfo {pages} {033086} (\bibinfo {year}
  {2019})}\BibitemShut {NoStop}%
\bibitem [{\citenamefont {Mark}\ and\ \citenamefont
  {Motrunich}(2020)}]{Mark2020b}%
  \BibitemOpen
  \bibfield  {author} {\bibinfo {author} {\bibfnamefont {D.~K.}\ \bibnamefont
  {Mark}}\ and\ \bibinfo {author} {\bibfnamefont {O.~I.}\ \bibnamefont
  {Motrunich}},\ }\bibfield  {title} {\bibinfo {title}
  {$\ensuremath{\eta}$-pairing states as true scars in an extended hubbard
  model},\ }\href {https://doi.org/10.1103/PhysRevB.102.075132} {\bibfield
  {journal} {\bibinfo  {journal} {Phys. Rev. B}\ }\textbf {\bibinfo {volume}
  {102}},\ \bibinfo {pages} {075132} (\bibinfo {year} {2020})}\BibitemShut
  {NoStop}%
\bibitem [{\citenamefont {Senko}\ \emph {et~al.}(2015)\citenamefont {Senko},
  \citenamefont {Richerme}, \citenamefont {Smith}, \citenamefont {Lee},
  \citenamefont {Cohen}, \citenamefont {Retzker},\ and\ \citenamefont
  {Monroe}}]{senko2015realization}%
  \BibitemOpen
  \bibfield  {author} {\bibinfo {author} {\bibfnamefont {C.}~\bibnamefont
  {Senko}}, \bibinfo {author} {\bibfnamefont {P.}~\bibnamefont {Richerme}},
  \bibinfo {author} {\bibfnamefont {J.}~\bibnamefont {Smith}}, \bibinfo
  {author} {\bibfnamefont {A.}~\bibnamefont {Lee}}, \bibinfo {author}
  {\bibfnamefont {I.}~\bibnamefont {Cohen}}, \bibinfo {author} {\bibfnamefont
  {A.}~\bibnamefont {Retzker}},\ and\ \bibinfo {author} {\bibfnamefont
  {C.}~\bibnamefont {Monroe}},\ }\bibfield  {title} {\bibinfo {title}
  {Realization of a quantum integer-spin chain with controllable
  interactions},\ }\href {https://doi.org/10.1103/PhysRevX.5.021026} {\bibfield
   {journal} {\bibinfo  {journal} {Phys. Rev. X}\ }\textbf {\bibinfo {volume}
  {5}},\ \bibinfo {pages} {021026} (\bibinfo {year} {2015})}\BibitemShut
  {NoStop}%
\bibitem [{\citenamefont {O'Dea}\ \emph {et~al.}(2020)\citenamefont {O'Dea},
  \citenamefont {Burnell}, \citenamefont {Chandran},\ and\ \citenamefont
  {Khemani}}]{Odea2020}%
  \BibitemOpen
  \bibfield  {author} {\bibinfo {author} {\bibfnamefont {N.}~\bibnamefont
  {O'Dea}}, \bibinfo {author} {\bibfnamefont {F.}~\bibnamefont {Burnell}},
  \bibinfo {author} {\bibfnamefont {A.}~\bibnamefont {Chandran}},\ and\
  \bibinfo {author} {\bibfnamefont {V.}~\bibnamefont {Khemani}},\ }\bibfield
  {title} {\bibinfo {title} {From tunnels to towers: Quantum scars from lie
  algebras and $q$-deformed lie algebras},\ }\href
  {https://doi.org/10.1103/PhysRevResearch.2.043305} {\bibfield  {journal}
  {\bibinfo  {journal} {Phys. Rev. Res.}\ }\textbf {\bibinfo {volume} {2}},\
  \bibinfo {pages} {043305} (\bibinfo {year} {2020})}\BibitemShut {NoStop}%
\bibitem [{\citenamefont {Bernien}\ \emph {et~al.}(2017)\citenamefont
  {Bernien}, \citenamefont {Schwartz}, \citenamefont {Keesling}, \citenamefont
  {Levine}, \citenamefont {Omran}, \citenamefont {Pichler}, \citenamefont
  {Choi}, \citenamefont {Zibrov}, \citenamefont {Endres}, \citenamefont
  {Greiner}, \citenamefont {Vuleti{\'c}},\ and\ \citenamefont
  {Lukin}}]{Lukin2017Rydberg}%
  \BibitemOpen
  \bibfield  {author} {\bibinfo {author} {\bibfnamefont {H.}~\bibnamefont
  {Bernien}}, \bibinfo {author} {\bibfnamefont {S.}~\bibnamefont {Schwartz}},
  \bibinfo {author} {\bibfnamefont {A.}~\bibnamefont {Keesling}}, \bibinfo
  {author} {\bibfnamefont {H.}~\bibnamefont {Levine}}, \bibinfo {author}
  {\bibfnamefont {A.}~\bibnamefont {Omran}}, \bibinfo {author} {\bibfnamefont
  {H.}~\bibnamefont {Pichler}}, \bibinfo {author} {\bibfnamefont
  {S.}~\bibnamefont {Choi}}, \bibinfo {author} {\bibfnamefont {A.~S.}\
  \bibnamefont {Zibrov}}, \bibinfo {author} {\bibfnamefont {M.}~\bibnamefont
  {Endres}}, \bibinfo {author} {\bibfnamefont {M.}~\bibnamefont {Greiner}},
  \bibinfo {author} {\bibfnamefont {V.}~\bibnamefont {Vuleti{\'c}}},\ and\
  \bibinfo {author} {\bibfnamefont {M.~D.}\ \bibnamefont {Lukin}},\ }\bibfield
  {title} {\bibinfo {title} {Probing many-body dynamics on a 51-atom quantum
  simulator},\ }\href {https://doi.org/10.1038/nature24622} {\bibfield
  {journal} {\bibinfo  {journal} {Nature}\ }\textbf {\bibinfo {volume} {551}},\
  \bibinfo {pages} {579} (\bibinfo {year} {2017})}\BibitemShut {NoStop}%
\bibitem [{\citenamefont {Barenco}\ \emph {et~al.}(1995)\citenamefont
  {Barenco}, \citenamefont {Bennett}, \citenamefont {Cleve}, \citenamefont
  {DiVincenzo}, \citenamefont {Margolus}, \citenamefont {Shor}, \citenamefont
  {Sleator}, \citenamefont {Smolin},\ and\ \citenamefont
  {Weinfurter}}]{Shor1995Elem}%
  \BibitemOpen
  \bibfield  {author} {\bibinfo {author} {\bibfnamefont {A.}~\bibnamefont
  {Barenco}}, \bibinfo {author} {\bibfnamefont {C.~H.}\ \bibnamefont
  {Bennett}}, \bibinfo {author} {\bibfnamefont {R.}~\bibnamefont {Cleve}},
  \bibinfo {author} {\bibfnamefont {D.~P.}\ \bibnamefont {DiVincenzo}},
  \bibinfo {author} {\bibfnamefont {N.}~\bibnamefont {Margolus}}, \bibinfo
  {author} {\bibfnamefont {P.}~\bibnamefont {Shor}}, \bibinfo {author}
  {\bibfnamefont {T.}~\bibnamefont {Sleator}}, \bibinfo {author} {\bibfnamefont
  {J.~A.}\ \bibnamefont {Smolin}},\ and\ \bibinfo {author} {\bibfnamefont
  {H.}~\bibnamefont {Weinfurter}},\ }\bibfield  {title} {\bibinfo {title}
  {Elementary gates for quantum computation},\ }\href
  {https://doi.org/10.1103/PhysRevA.52.3457} {\bibfield  {journal} {\bibinfo
  {journal} {Phys. Rev. A}\ }\textbf {\bibinfo {volume} {52}},\ \bibinfo
  {pages} {3457} (\bibinfo {year} {1995})}\BibitemShut {NoStop}%
\bibitem [{\citenamefont {Shi}(2003)}]{shi2002both}%
  \BibitemOpen
  \bibfield  {author} {\bibinfo {author} {\bibfnamefont {Y.}~\bibnamefont
  {Shi}},\ }\bibfield  {title} {\bibinfo {title} {Both toffoli and
  controlled-not need little help to do universal quantum computing},\ }\href
  {https://dl.acm.org/doi/10.5555/2011508.2011515} {\bibfield  {journal}
  {\bibinfo  {journal} {Quantum Info. Comput.}\ }\textbf {\bibinfo {volume}
  {3}},\ \bibinfo {pages} {84–92} (\bibinfo {year} {2003})}\BibitemShut
  {NoStop}%
\bibitem [{\citenamefont {Shende}\ and\ \citenamefont
  {Markov}(2009)}]{Shende2009Toffoli}%
  \BibitemOpen
  \bibfield  {author} {\bibinfo {author} {\bibfnamefont {V.~V.}\ \bibnamefont
  {Shende}}\ and\ \bibinfo {author} {\bibfnamefont {I.~L.}\ \bibnamefont
  {Markov}},\ }\bibfield  {title} {\bibinfo {title} {On the cnot-cost of
  toffoli gates},\ }\href {https://dl.acm.org/doi/abs/10.5555/2011791.2011799}
  {\bibfield  {journal} {\bibinfo  {journal} {Quantum Info. Comput.}\ }\textbf
  {\bibinfo {volume} {9}},\ \bibinfo {pages} {461–486} (\bibinfo {year}
  {2009})}\BibitemShut {NoStop}%
\bibitem [{\citenamefont {Wu}\ \emph {et~al.}(2025)\citenamefont {Wu},
  \citenamefont {Yang}, \citenamefont {Claeys},\ and\ \citenamefont
  {Zhao}}]{wu2025engineering}%
  \BibitemOpen
  \bibfield  {author} {\bibinfo {author} {\bibfnamefont {R.}~\bibnamefont
  {Wu}}, \bibinfo {author} {\bibfnamefont {B.}~\bibnamefont {Yang}}, \bibinfo
  {author} {\bibfnamefont {P.~W.}\ \bibnamefont {Claeys}},\ and\ \bibinfo
  {author} {\bibfnamefont {H.}~\bibnamefont {Zhao}},\ }\bibfield  {title}
  {\bibinfo {title} {Engineering long-range and multi-body interactions via
  global kinetic constraints},\ }\href {https://arxiv.org/abs/2505.08390}
  {\bibfield  {journal} {\bibinfo  {journal} {arXiv preprint arXiv:2505.08390}\
  } (\bibinfo {year} {2025})}\BibitemShut {NoStop}%
\bibitem [{\citenamefont {Monz}\ \emph {et~al.}(2009)\citenamefont {Monz},
  \citenamefont {Kim}, \citenamefont {H\"ansel}, \citenamefont {Riebe},
  \citenamefont {Villar}, \citenamefont {Schindler}, \citenamefont {Chwalla},
  \citenamefont {Hennrich},\ and\ \citenamefont
  {Blatt}}]{Monz2009RealizationToffoli}%
  \BibitemOpen
  \bibfield  {author} {\bibinfo {author} {\bibfnamefont {T.}~\bibnamefont
  {Monz}}, \bibinfo {author} {\bibfnamefont {K.}~\bibnamefont {Kim}}, \bibinfo
  {author} {\bibfnamefont {W.}~\bibnamefont {H\"ansel}}, \bibinfo {author}
  {\bibfnamefont {M.}~\bibnamefont {Riebe}}, \bibinfo {author} {\bibfnamefont
  {A.~S.}\ \bibnamefont {Villar}}, \bibinfo {author} {\bibfnamefont
  {P.}~\bibnamefont {Schindler}}, \bibinfo {author} {\bibfnamefont
  {M.}~\bibnamefont {Chwalla}}, \bibinfo {author} {\bibfnamefont
  {M.}~\bibnamefont {Hennrich}},\ and\ \bibinfo {author} {\bibfnamefont
  {R.}~\bibnamefont {Blatt}},\ }\bibfield  {title} {\bibinfo {title}
  {Realization of the quantum toffoli gate with trapped ions},\ }\href
  {https://doi.org/10.1103/PhysRevLett.102.040501} {\bibfield  {journal}
  {\bibinfo  {journal} {Phys. Rev. Lett.}\ }\textbf {\bibinfo {volume} {102}},\
  \bibinfo {pages} {040501} (\bibinfo {year} {2009})}\BibitemShut {NoStop}%
\bibitem [{\citenamefont {Evered}\ \emph {et~al.}(2023)\citenamefont {Evered},
  \citenamefont {Bluvstein}, \citenamefont {Kalinowski}, \citenamefont {Ebadi},
  \citenamefont {Manovitz}, \citenamefont {Zhou}, \citenamefont {Li},
  \citenamefont {Geim}, \citenamefont {Wang}, \citenamefont {Maskara},
  \citenamefont {Levine}, \citenamefont {Semeghini}, \citenamefont {Greiner},
  \citenamefont {Vuletić},\ and\ \citenamefont {Lukin}}]{Evered2023HighFid}%
  \BibitemOpen
  \bibfield  {author} {\bibinfo {author} {\bibfnamefont {S.~J.}\ \bibnamefont
  {Evered}}, \bibinfo {author} {\bibfnamefont {D.}~\bibnamefont {Bluvstein}},
  \bibinfo {author} {\bibfnamefont {M.}~\bibnamefont {Kalinowski}}, \bibinfo
  {author} {\bibfnamefont {S.}~\bibnamefont {Ebadi}}, \bibinfo {author}
  {\bibfnamefont {T.}~\bibnamefont {Manovitz}}, \bibinfo {author}
  {\bibfnamefont {H.}~\bibnamefont {Zhou}}, \bibinfo {author} {\bibfnamefont
  {S.~H.}\ \bibnamefont {Li}}, \bibinfo {author} {\bibfnamefont {A.~A.}\
  \bibnamefont {Geim}}, \bibinfo {author} {\bibfnamefont {T.~T.}\ \bibnamefont
  {Wang}}, \bibinfo {author} {\bibfnamefont {N.}~\bibnamefont {Maskara}},
  \bibinfo {author} {\bibfnamefont {H.}~\bibnamefont {Levine}}, \bibinfo
  {author} {\bibfnamefont {G.}~\bibnamefont {Semeghini}}, \bibinfo {author}
  {\bibfnamefont {M.}~\bibnamefont {Greiner}}, \bibinfo {author} {\bibfnamefont
  {V.}~\bibnamefont {Vuletić}},\ and\ \bibinfo {author} {\bibfnamefont
  {M.~D.}\ \bibnamefont {Lukin}},\ }\bibfield  {title} {\bibinfo {title}
  {High-fidelity parallel entangling gates on a neutral-atom quantum
  computer},\ }\href {https://doi.org/10.1038/s41586-023-06481-y} {\bibfield
  {journal} {\bibinfo  {journal} {Nature}\ }\textbf {\bibinfo {volume} {622}},\
  \bibinfo {pages} {268} (\bibinfo {year} {2023})}\BibitemShut {NoStop}%
\bibitem [{\citenamefont {Cao}\ \emph {et~al.}(2024)\citenamefont {Cao},
  \citenamefont {Eckner}, \citenamefont {Lukin~Yelin}, \citenamefont {Young},
  \citenamefont {Jandura}, \citenamefont {Yan}, \citenamefont {Kim},
  \citenamefont {Pupillo}, \citenamefont {Ye}, \citenamefont {Darkwah~Oppong}
  \emph {et~al.}}]{cao2024multi}%
  \BibitemOpen
  \bibfield  {author} {\bibinfo {author} {\bibfnamefont {A.}~\bibnamefont
  {Cao}}, \bibinfo {author} {\bibfnamefont {W.~J.}\ \bibnamefont {Eckner}},
  \bibinfo {author} {\bibfnamefont {T.}~\bibnamefont {Lukin~Yelin}}, \bibinfo
  {author} {\bibfnamefont {A.~W.}\ \bibnamefont {Young}}, \bibinfo {author}
  {\bibfnamefont {S.}~\bibnamefont {Jandura}}, \bibinfo {author} {\bibfnamefont
  {L.}~\bibnamefont {Yan}}, \bibinfo {author} {\bibfnamefont {K.}~\bibnamefont
  {Kim}}, \bibinfo {author} {\bibfnamefont {G.}~\bibnamefont {Pupillo}},
  \bibinfo {author} {\bibfnamefont {J.}~\bibnamefont {Ye}}, \bibinfo {author}
  {\bibfnamefont {N.}~\bibnamefont {Darkwah~Oppong}}, \emph {et~al.},\
  }\bibfield  {title} {\bibinfo {title} {Multi-qubit gates and schr{\"o}dinger
  cat states in an optical clock},\ }\href
  {https://doi.org/10.1038/s41586-024-07913-z} {\bibfield  {journal} {\bibinfo
  {journal} {Nature}\ }\textbf {\bibinfo {volume} {634}},\ \bibinfo {pages}
  {315} (\bibinfo {year} {2024})}\BibitemShut {NoStop}%
\bibitem [{\citenamefont {Nie}\ \emph {et~al.}(2024)\citenamefont {Nie},
  \citenamefont {Zi},\ and\ \citenamefont {Sun}}]{nie2024quantum}%
  \BibitemOpen
  \bibfield  {author} {\bibinfo {author} {\bibfnamefont {J.}~\bibnamefont
  {Nie}}, \bibinfo {author} {\bibfnamefont {W.}~\bibnamefont {Zi}},\ and\
  \bibinfo {author} {\bibfnamefont {X.}~\bibnamefont {Sun}},\ }\bibfield
  {title} {\bibinfo {title} {Quantum circuit for multi-qubit toffoli gate with
  optimal resource},\ }\href {https://arxiv.org/abs/2402.05053} {\bibfield
  {journal} {\bibinfo  {journal} {arXiv preprint arXiv:2402.05053}\ } (\bibinfo
  {year} {2024})}\BibitemShut {NoStop}%
\bibitem [{\citenamefont {Bao}\ \emph {et~al.}(2024)\citenamefont {Bao},
  \citenamefont {Xu}, \citenamefont {Song}, \citenamefont {Wang}, \citenamefont
  {Xiang}, \citenamefont {Zhu}, \citenamefont {Chen}, \citenamefont {Jin},
  \citenamefont {Zhu}, \citenamefont {Gao} \emph {et~al.}}]{Huangbiao2024nc}%
  \BibitemOpen
  \bibfield  {author} {\bibinfo {author} {\bibfnamefont {Z.}~\bibnamefont
  {Bao}}, \bibinfo {author} {\bibfnamefont {S.}~\bibnamefont {Xu}}, \bibinfo
  {author} {\bibfnamefont {Z.}~\bibnamefont {Song}}, \bibinfo {author}
  {\bibfnamefont {K.}~\bibnamefont {Wang}}, \bibinfo {author} {\bibfnamefont
  {L.}~\bibnamefont {Xiang}}, \bibinfo {author} {\bibfnamefont
  {Z.}~\bibnamefont {Zhu}}, \bibinfo {author} {\bibfnamefont {J.}~\bibnamefont
  {Chen}}, \bibinfo {author} {\bibfnamefont {F.}~\bibnamefont {Jin}}, \bibinfo
  {author} {\bibfnamefont {X.}~\bibnamefont {Zhu}}, \bibinfo {author}
  {\bibfnamefont {Y.}~\bibnamefont {Gao}}, \emph {et~al.},\ }\bibfield  {title}
  {\bibinfo {title} {Creating and controlling global
  greenberger--horne--zeilinger entanglement on quantum processors},\ }\href
  {https://doi.org/10.1038/s41467-024-53140-5} {\bibfield  {journal} {\bibinfo
  {journal} {Nature Communications}\ }\textbf {\bibinfo {volume} {15}},\
  \bibinfo {pages} {8823} (\bibinfo {year} {2024})}\BibitemShut {NoStop}%
\bibitem [{\citenamefont {Iadecola}\ \emph {et~al.}(2025)\citenamefont
  {Iadecola}, \citenamefont {Wilson},\ and\ \citenamefont
  {Pixley}}]{Thomas2025PRXQ}%
  \BibitemOpen
  \bibfield  {author} {\bibinfo {author} {\bibfnamefont {T.}~\bibnamefont
  {Iadecola}}, \bibinfo {author} {\bibfnamefont {J.~H.}\ \bibnamefont
  {Wilson}},\ and\ \bibinfo {author} {\bibfnamefont {J.}~\bibnamefont
  {Pixley}},\ }\bibfield  {title} {\bibinfo {title} {Concomitant entanglement
  and control criticality driven by collective measurements},\ }\href
  {https://doi.org/10.1103/PRXQuantum.6.010351} {\bibfield  {journal} {\bibinfo
   {journal} {PRX Quantum}\ }\textbf {\bibinfo {volume} {6}},\ \bibinfo {pages}
  {010351} (\bibinfo {year} {2025})}\BibitemShut {NoStop}%
\bibitem [{\citenamefont {Zhao}\ \emph
  {et~al.}(2021{\natexlab{a}})\citenamefont {Zhao}, \citenamefont {Smith},
  \citenamefont {Mintert},\ and\ \citenamefont {Knolle}}]{Zhao2021Orthogonal}%
  \BibitemOpen
  \bibfield  {author} {\bibinfo {author} {\bibfnamefont {H.}~\bibnamefont
  {Zhao}}, \bibinfo {author} {\bibfnamefont {A.}~\bibnamefont {Smith}},
  \bibinfo {author} {\bibfnamefont {F.}~\bibnamefont {Mintert}},\ and\ \bibinfo
  {author} {\bibfnamefont {J.}~\bibnamefont {Knolle}},\ }\bibfield  {title}
  {\bibinfo {title} {Orthogonal quantum many-body scars},\ }\href
  {https://doi.org/10.1103/PhysRevLett.127.150601} {\bibfield  {journal}
  {\bibinfo  {journal} {Phys. Rev. Lett.}\ }\textbf {\bibinfo {volume} {127}},\
  \bibinfo {pages} {150601} (\bibinfo {year} {2021}{\natexlab{a}})}\BibitemShut
  {NoStop}%
\bibitem [{\citenamefont {Langlett}\ \emph {et~al.}(2022)\citenamefont
  {Langlett}, \citenamefont {Yang}, \citenamefont {Wildeboer}, \citenamefont
  {Gorshkov}, \citenamefont {Iadecola},\ and\ \citenamefont
  {Xu}}]{Langlett2022volume}%
  \BibitemOpen
  \bibfield  {author} {\bibinfo {author} {\bibfnamefont {C.~M.}\ \bibnamefont
  {Langlett}}, \bibinfo {author} {\bibfnamefont {Z.-C.}\ \bibnamefont {Yang}},
  \bibinfo {author} {\bibfnamefont {J.}~\bibnamefont {Wildeboer}}, \bibinfo
  {author} {\bibfnamefont {A.~V.}\ \bibnamefont {Gorshkov}}, \bibinfo {author}
  {\bibfnamefont {T.}~\bibnamefont {Iadecola}},\ and\ \bibinfo {author}
  {\bibfnamefont {S.}~\bibnamefont {Xu}},\ }\bibfield  {title} {\bibinfo
  {title} {Rainbow scars: From area to volume law},\ }\href
  {https://doi.org/10.1103/PhysRevB.105.L060301} {\bibfield  {journal}
  {\bibinfo  {journal} {Phys. Rev. B}\ }\textbf {\bibinfo {volume} {105}},\
  \bibinfo {pages} {L060301} (\bibinfo {year} {2022})}\BibitemShut {NoStop}%
\bibitem [{\citenamefont {Mukherjee}\ \emph {et~al.}(2025)\citenamefont
  {Mukherjee}, \citenamefont {Turner}, \citenamefont {Szyniszewski},\ and\
  \citenamefont {Pal}}]{mukherjee2025symmetrictensorscarstunable}%
  \BibitemOpen
  \bibfield  {author} {\bibinfo {author} {\bibfnamefont {B.}~\bibnamefont
  {Mukherjee}}, \bibinfo {author} {\bibfnamefont {C.~J.}\ \bibnamefont
  {Turner}}, \bibinfo {author} {\bibfnamefont {M.}~\bibnamefont
  {Szyniszewski}},\ and\ \bibinfo {author} {\bibfnamefont {A.}~\bibnamefont
  {Pal}},\ }\bibfield  {title} {\bibinfo {title} {Symmetric tensor scars with
  tunable entanglement from volume to area law},\ }\href
  {https://arxiv.org/abs/2501.14024} {\bibfield  {journal} {\bibinfo  {journal}
  {arXiv preprint arXiv:2501.14024}\ } (\bibinfo {year} {2025})}\BibitemShut
  {NoStop}%
\bibitem [{\citenamefont {Ho}\ \emph {et~al.}(2023)\citenamefont {Ho},
  \citenamefont {Mori}, \citenamefont {Abanin},\ and\ \citenamefont {{Dalla
  Torre}}}]{ho2023quantum}%
  \BibitemOpen
  \bibfield  {author} {\bibinfo {author} {\bibfnamefont {W.~W.}\ \bibnamefont
  {Ho}}, \bibinfo {author} {\bibfnamefont {T.}~\bibnamefont {Mori}}, \bibinfo
  {author} {\bibfnamefont {D.~A.}\ \bibnamefont {Abanin}},\ and\ \bibinfo
  {author} {\bibfnamefont {E.~G.}\ \bibnamefont {{Dalla Torre}}},\ }\bibfield
  {title} {\bibinfo {title} {Quantum and classical floquet prethermalization},\
  }\href {https://doi.org/https://doi.org/10.1016/j.aop.2023.169297} {\bibfield
   {journal} {\bibinfo  {journal} {Annals of Physics}\ }\textbf {\bibinfo
  {volume} {454}},\ \bibinfo {pages} {169297} (\bibinfo {year}
  {2023})}\BibitemShut {NoStop}%
\bibitem [{\citenamefont {Zhao}\ \emph
  {et~al.}(2021{\natexlab{b}})\citenamefont {Zhao}, \citenamefont {Mintert},
  \citenamefont {Moessner},\ and\ \citenamefont {Knolle}}]{zhao2021random}%
  \BibitemOpen
  \bibfield  {author} {\bibinfo {author} {\bibfnamefont {H.}~\bibnamefont
  {Zhao}}, \bibinfo {author} {\bibfnamefont {F.}~\bibnamefont {Mintert}},
  \bibinfo {author} {\bibfnamefont {R.}~\bibnamefont {Moessner}},\ and\
  \bibinfo {author} {\bibfnamefont {J.}~\bibnamefont {Knolle}},\ }\bibfield
  {title} {\bibinfo {title} {Random multipolar driving: Tunably slow heating
  through spectral engineering},\ }\href
  {https://doi.org/10.1103/PhysRevLett.126.040601} {\bibfield  {journal}
  {\bibinfo  {journal} {Phys. Rev. Lett.}\ }\textbf {\bibinfo {volume} {126}},\
  \bibinfo {pages} {040601} (\bibinfo {year} {2021}{\natexlab{b}})}\BibitemShut
  {NoStop}%
\bibitem [{\citenamefont {Pilatowsky-Cameo}\ \emph {et~al.}(2025)\citenamefont
  {Pilatowsky-Cameo}, \citenamefont {Choi},\ and\ \citenamefont
  {Ho}}]{pilatowsky2025critically}%
  \BibitemOpen
  \bibfield  {author} {\bibinfo {author} {\bibfnamefont {S.}~\bibnamefont
  {Pilatowsky-Cameo}}, \bibinfo {author} {\bibfnamefont {S.}~\bibnamefont
  {Choi}},\ and\ \bibinfo {author} {\bibfnamefont {W.~W.}\ \bibnamefont {Ho}},\
  }\bibfield  {title} {\bibinfo {title} {Critically slow hilbert-space
  ergodicity in quantum morphic drives},\ }\href
  {https://arxiv.org/abs/2502.06936} {\bibfield  {journal} {\bibinfo  {journal}
  {arXiv preprint arXiv:2502.06936}\ } (\bibinfo {year} {2025})}\BibitemShut
  {NoStop}%
\bibitem [{\citenamefont {Pocklington}\ and\ \citenamefont
  {Clerk}(2025)}]{pocklington2025accelerating}%
  \BibitemOpen
  \bibfield  {author} {\bibinfo {author} {\bibfnamefont {A.}~\bibnamefont
  {Pocklington}}\ and\ \bibinfo {author} {\bibfnamefont {A.~A.}\ \bibnamefont
  {Clerk}},\ }\bibfield  {title} {\bibinfo {title} {Accelerating dissipative
  state preparation with adaptive open quantum dynamics},\ }\href
  {https://doi.org/10.1103/PhysRevLett.134.050603} {\bibfield  {journal}
  {\bibinfo  {journal} {Phys. Rev. Lett.}\ }\textbf {\bibinfo {volume} {134}},\
  \bibinfo {pages} {050603} (\bibinfo {year} {2025})}\BibitemShut {NoStop}%
\bibitem [{\citenamefont {Sinkovicz}\ \emph {et~al.}(2015)\citenamefont
  {Sinkovicz}, \citenamefont {Kurucz}, \citenamefont {Kiss},\ and\
  \citenamefont {Asb\'oth}}]{Sinkovicz2015}%
  \BibitemOpen
  \bibfield  {author} {\bibinfo {author} {\bibfnamefont {P.}~\bibnamefont
  {Sinkovicz}}, \bibinfo {author} {\bibfnamefont {Z.}~\bibnamefont {Kurucz}},
  \bibinfo {author} {\bibfnamefont {T.}~\bibnamefont {Kiss}},\ and\ \bibinfo
  {author} {\bibfnamefont {J.~K.}\ \bibnamefont {Asb\'oth}},\ }\bibfield
  {title} {\bibinfo {title} {Quantized recurrence time in unital iterated open
  quantum dynamics},\ }\href {https://doi.org/10.1103/PhysRevA.91.042108}
  {\bibfield  {journal} {\bibinfo  {journal} {Phys. Rev. A}\ }\textbf {\bibinfo
  {volume} {91}},\ \bibinfo {pages} {042108} (\bibinfo {year}
  {2015})}\BibitemShut {NoStop}%
\end{thebibliography}
%

\clearpage

\noindent {\bf \large{End matter}}
~\\

\begin{figure}[h]
\centering
\includegraphics[width=0.95\linewidth]{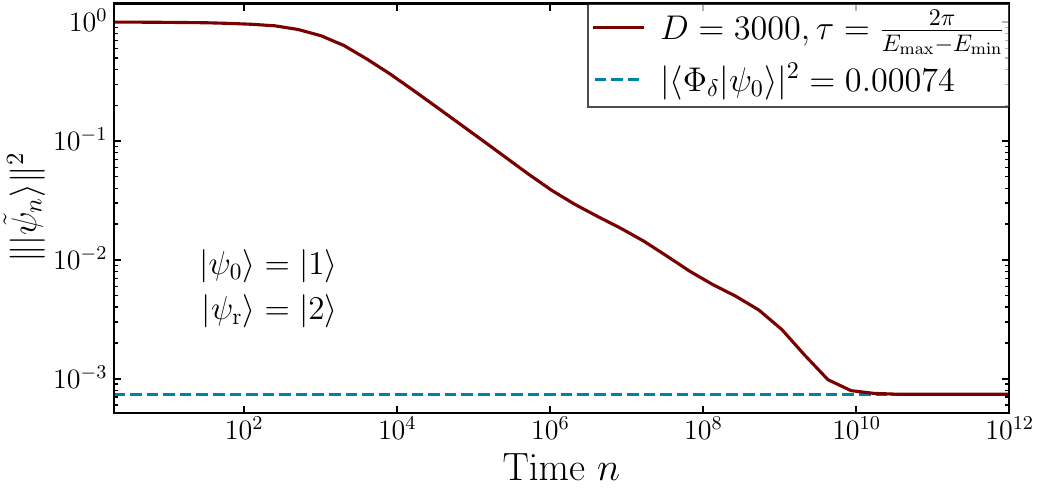}
\caption{The system is filtered to the target superposition (blue dashed line) at long times. The unitary evolution of the system is governed by a random matrix. 
} 
\label{fig:randmat}
\end{figure}

{\em Dark states in random matrices.---}
Many-body interacting systems exhibit spectral and dynamical features
that can be well described by random matrix theory.
To show that our state filtration protocol applies to generic many-body systems,
we employ a $D$-dimensional random matrix drawn from the Gaussian Orthogonal Ensemble (GOE).
Specifically, we sample a real symmetric matrix whose elements are independently drawn from Gaussian distributions: diagonal entries with zero mean and variance $2/D$, and off-diagonal entries with zero mean and variance $1/D$.
This scaling ensures a bounded eigenvalue spectrum in the large-$D$ limit.
We represent this matrix in the computational basis $\{\ket{j}\}$, where $j$ runs from $1$ to $D$.
The removal state $\ket{\psi_{\rm r}}$ is chosen as $\ket{2}$.
and the initial state is $\ket{\psi_0} = \ket{1}$.
The measurement period $\tau$ is chosen to satisfy the resonance condition, $\exp(-i E_{\rm max} \tau) = \exp(-i E_{\rm min} \tau)$,
where $E_{\rm max}$ ($E_{\rm min}$)
is the maximum (minimum) of the eigenvalues of the random matrix. This leads to the dark state as a cat state composed of eigenstates with distinct energies,
$\ket{\Phi_\delta} =
{\rm N}(\langle \psi_{\rm r} | E_{\rm min} \rangle | E_{\rm max} \rangle 
- \langle \psi_{\rm r} | E_{\rm max} \rangle | E_{\rm min} \rangle)$.
With long-time filtration,
the amplitude of the initial state projected onto the dark state persists,
converging to
$\||\tilde{\psi}_n \rangle \|^2 = \|({\cal F}^\tau_{\psi_{\rm r}})^n \ket{\psi_0}\|^2
\to |\langle\Phi_\delta | \psi_0 \rangle|^2$ for large $n$
(see Eq. (4) in the main text). The numerical result is presented in Fig. \ref{fig:randmat}, where $\||\tilde{\psi}_n \rangle \|^2$ converges to the target (blue dashed line) at long times.
This suggests that our dark-state engineering and state filtration protocol
are applicable to generic many-body systems,
enabling the preparation of controllable superpositions over energy eigenstates
via non-local measurement.

\begin{widetext}

{\em Other target states.---}
The two target states in the spin-1 XY model shown in the main text are representative examples. Here we provide more possibilities by tuning $h\tau$ to other resonance conditions.
These cases include multifold degeneracy of $U(\tau)$, i.e., the degeneracy $g>2$,
inducing $g-1$ dark states and a rich variety of targets. 
In general, resonances occur at $h\tau = \pi p/q$,
where $p$ and $q$ are integers satisfying $p \geq 1$ and $1 \leq q \leq L$.
However, since the system fully revives (returning to the initial state)
whenever $h\tau = k\pi$ for integer $k$,
all distinct resonances are contained within the fundamental interval $h\tau \in [0,\pi]$.
As the number of resonances increases with $L$,
for simplicity,
we explicitly list the dark states and the resulting superpositions for $L=6$, see the summary in Table \ref{tab:darkstates}.

For $g>2$, $g-1$ dark states can be recursively constructed with the Gram-Schmidt process
using the following determinant~\cite{Thiel2020D},
\begin{equation}\label{eq:determinantDark}
    |\Phi_{ \delta}\rangle
    =
    {\rm N_\delta}\begin{vmatrix}
    |{\cal C}_{1}\rangle&|{\cal C}_{2}\rangle&\cdots&|{\cal C}_{\delta+1}\rangle \\
    \langle\psi_{\rm r}|{\cal C}_{1}\rangle & \langle\psi_{\rm r}|{\cal C}_{2}\rangle & \cdots & \langle\psi_{\rm r}|{\cal C}_{\delta+1}\rangle \\ 
    \langle\Phi_1|{\cal C}_{1}\rangle & \langle\Phi_1|{\cal C}_{2}\rangle & \cdots & \langle\Phi_1|{\cal C}_{\delta+1}\rangle \\ 
    \vdots & \vdots & \ddots & \vdots \\
    \langle\Phi_{\delta-1}|{\cal C}_{1}\rangle & \langle\Phi_{\delta-1}|{\cal C}_{2}\rangle & \cdots & \langle\Phi_{\delta-1}|{\cal C}_{\delta+1} \rangle
    \end{vmatrix}.
\end{equation}
Here $\delta \le g-1$,
and $N_\delta$ is the normalization factor for the $\delta$th dark state $\ket{\Phi_{\delta}}$.
The set $\{ |{\cal C}_{1}\rangle, |{\cal C}_{2}\rangle, \dots, |{\cal C}_{g}\rangle \}$ denotes the degenerate eigenstates.
For example,
when $h\tau = \pi/3$, and for $L=6$,
we have a $3$-fold degeneracy,
$\exp(-i E_0\tau)=\exp(-i E_3\tau)=\exp(-i E_6\tau)$ (recall that $E_n = (2n-L)h + DL$).
Hence, in this ($g=3$)-dimensional subspace there are two dark states emerging, which can be
computed recursively,
\begin{equation}\label{eq:determinant1}
\begin{aligned}
    |\Phi_{1}\rangle
    &=
    {\rm N}_1
    \begin{vmatrix}
    |{\cal B}_{0}\rangle & |{\cal B}_{3}\rangle \\
    \langle\psi_{\rm r}|{\cal B}_{0}\rangle & \langle\psi_{\rm r}|{\cal B}_{3}\rangle
    \end{vmatrix} 
    = {\rm N}_1 (\langle\psi_{\rm r}|{\cal B}_{3}\rangle |{\cal B}_{0}\rangle 
    - \langle\psi_{\rm r}|{\cal B}_{0}\rangle |{\cal B}_{3}\rangle),
\end{aligned}
\end{equation}
and
\begin{equation}\label{eq:determinant2}
    |\Phi_{2}\rangle
    =
    {\rm N_2}\begin{vmatrix}
    |{\cal B}_{0}\rangle&|{\cal B}_{3}\rangle&|{\cal B}_{6}\rangle \\
    \langle\psi_{\rm r}|{\cal B}_{0}\rangle & \langle\psi_{\rm r}|{\cal B}_{3}\rangle &  \langle\psi_{\rm r}|{\cal B}_{6}\rangle \\ 
    \langle\Phi_1|{\cal B}_{0}\rangle & \langle\Phi_1|{\cal B}_3\rangle & \langle\Phi_1|{\cal B}_{6}\rangle 
    \end{vmatrix}.
\end{equation}

The dark states $|\Phi_{\delta}\rangle$ generated with Eq.~\eqref{eq:determinantDark}
hold the following properties:
(i) They are eigenstates of the unitary operator $U(\tau)$.
(ii) $|\psi_{\rm r}\rangle$ is orthogonal to the removal state, i.e. $\langle\psi_{\rm r}|\Phi_{\delta}\rangle$=0.
(iii) $\langle \Phi_{\delta}|\Phi_{\delta^\prime}\rangle=0$ if $\delta^\prime \neq \delta$.
Hence, they form a complete basis of the dark subspace and survive the filtration at stroboscopic times $t=n\tau$.
All the target states in Table \ref{tab:darkstates} can be explicitly obtained using the formula Eq.~\eqref{eq:determinantDark}.
For clarity, we label each dark state with a superscript
indicating the bi-magnon indices of its constituent eigenstates.
For example,
$\ket{\Phi^{(0,6)}} = {\rm N}(\langle \psi_{\rm r}|{\cal B}_6\rangle |{\cal B}_0\rangle
- \langle \psi_{\rm r}|{\cal B}_0\rangle |{\cal B}_6\rangle)$.
When multiple ($g>2$) eigenstates become resonant,
creating more than one dark state, we additionally use subscripts to distinguish among them. 
{Therefore, our protocol is flexible in preparing a variety of target superpositions with tunable properties.}
\begin{table*}[htbp]
    \centering
    \caption{Dark states and target states for different values of $h\tau$ and $L=6$.
    For clarity, we use the superscript to indicate the bi-magnon eigenstates
    composing the dark state,
    and the subscript to label the dark states when more than one dark states
    arising from a multifold resonance.
    $\eta$ denotes the overlap between one specific dark state and the initial state,
    e.g., $\eta^{(0,5)}= \langle \Phi^{(0,5)} | \psi_0 \rangle$.}
    \label{tab:darkstates}
    \renewcommand{\arraystretch}{1.25}
    \setlength{\tabcolsep}{7pt}
    \begin{tabular}{c c c}
        \hline\hline
        $h\tau$ & Dark States $\ket{\Phi}$ & Target States $\ket{\Psi_{\rm tar}}$ (Non-normalized) \\
        \hline
        $\frac{\pi}{6}, \frac{5\pi}{6}$ &
        \makecell{
            $\ket*{\Phi^{(0,6)}}$ 
        } &
        \makecell{
            $\ket*{\Phi^{(0,6)}} 
            $
        } \\ 
        [0.2ex]
        $\frac{\pi}{5}, \frac{4\pi}{5}$ &
        \makecell{
            $\ket*{\Phi^{(0,5)}}, \ket*{\Phi^{(1,6)}}$
        } &
        \makecell{
            $\eta^{(0,5)}|\Phi^{(0,5)}\rangle e^{-in\frac{4\pi}{5}} 
            + \eta^{(1,6)}|\Phi^{(1,6)}\rangle e^{-in\frac{6\pi}{5}}$
        } \\ 
        $\frac{\pi}{4}, \frac{3\pi}{4}$ &
        \makecell{
            $\ket*{\Phi^{(0,4)}}, \ket*{\Phi^{(1,5)}}, \ket*{\Phi^{(2,6)}}$
        } &
        \makecell{
            $\eta^{(0,4)}|\Phi^{(0,4)}\rangle e^{-in\frac{\pi}{2}}
            +\eta^{(1,5)}|\Phi^{(1,5)}\rangle e^{in\pi}
            +\eta^{(2,6)}|\Phi^{(2,6)}\rangle e^{-in\frac{3\pi}{2}}$
        } \\ [0.65ex]
        $\frac{\pi}{3}, \frac{2\pi}{3}$ &
        \makecell{
            $\ket*{\Phi^{(0,3,6)}_1},\ket*{\Phi^{(0,3,6)}_2}$,
            \\ [0.3ex]
            $\ket*{\Phi^{(1,4)}}, \ket*{\Phi^{(2,5)}}$
        } &
        \makecell{
            $\big[
            \eta_1^{(0,3,6)}|\Phi_1^{(0,3,6)}\rangle 
            +\eta_2^{(0,3,6)}|\Phi_2^{(0,3,6)}\rangle 
            \big] e^0 $ 
            \\
            $+ \eta^{(1,4)}|\Phi^{(1,4)}\rangle e^{-in\frac{2\pi}{3}}
            + \eta^{(2,5)}|\Phi^{(2,5)}\rangle e^{-in\frac{4\pi}{3}}$
        } \\ [0.75ex]
        $\frac{2\pi}{5}, \frac{3\pi}{5}$ &
        \makecell{
            $\ket*{\Phi^{(0,5)}}, \ket*{\Phi^{(1,6)}}$
        } &
        \makecell{
            $\eta^{(0,5)}|\Phi^{(0,5)}\rangle e^{in\frac{2\pi}{5}}
            +\eta^{(1,6)}|\Phi^{(1,6)}\rangle e^{-in\frac{2\pi}{5}}$
        } \\ [0.7ex]
        $\frac{\pi}{2}$ &
        \makecell{
            $\ket*{\Phi^{(0,2,4,6)}_1},\ket*{\Phi^{(0,2,4,6)}_2},\ket*{\Phi^{(0,2,4,6)}_3}$, \\ [0.3ex]
            $\ket*{\Phi^{(1,3,5)}_1}, \ket*{\Phi^{(1,3,5)}_2}$
        } &
        \makecell{
            $\big[ 
            \eta^{(0,2,4,6)}_1|\Phi^{(0,2,4,6)}_1\rangle 
            +\eta^{(0,2,4,6)}_2|\Phi^{(0,2,4,6)}_2\rangle
            +\eta^{(0,2,4,6)}_3|\Phi^{(0,2,4,6)}_3\rangle 
            \big] e^{in\pi}$ \\
            $+ 
            \big[ 
            \eta^{(1,3,5)}_1|\Phi^{(1,3,5)}_1\rangle + \eta^{(1,3,5)}_2|\Phi^{(1,3,5)}_2\rangle 
            \big] e^0$
        } \\
        \hline\hline
    \end{tabular}
\end{table*}

{\em Different scaling of the filtration time versus the system size.---}
As shown in the main text,
the two targets we choose require an exponential filtration time in $L$.
However, here we show that the other scaling behaviors can exist. Interestingly, some target states in Table \ref{tab:darkstates} demand a filtration time that decreases for larger system sizes.
This can be readily shown with the modulus of the dominant bright eigenvalue {(refer to the SM Note 1 for calculation details),}
see Fig.~\ref{fig:domBriEv} for the example of $h\tau=\pi/4$.
In this case,
the equally spaced energy levels of the spin-1 XY model
yield a $\pi/2$ angular difference between consecutive energy phases $e^{-i E_n \tau}$. Consequently, there are always four resonances,
each exhibiting a degeneracy that increases with system size $L$,
thus enriching the structure of the target states.
Similar behaviors can be witnessed also for other choices of $h\tau$,
say $h\tau=\pi/3, 2\pi/3, 3\pi/4,$ etc.,
and they are not exceptional cases, essentially attributed to the equally spaced energies. 
{As shown in Fig.~\ref{fig:domBriEv},
all bright eigenvalues go to $0$ while we increase the system size. Hence, the target state can be immediately created after the first measurement.}
\begin{figure*}[h]
\centering
\includegraphics[width=0.45\linewidth]{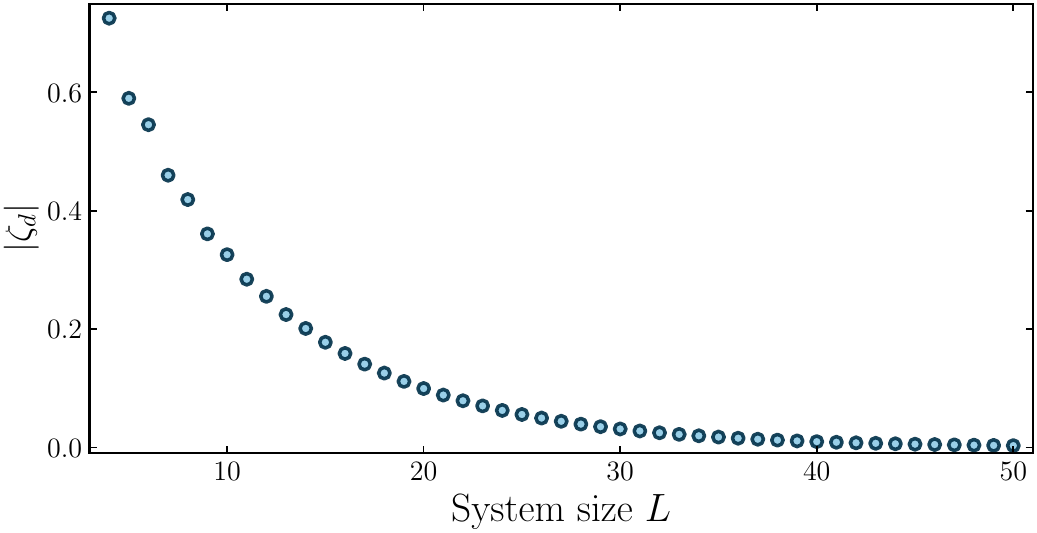}
\caption{The modulus of the dominant bright eigenvalue decreases as the system size grows.
This indicates that the preparation time for the target (c.f. $\ket*{\Psi_{\rm tar}}$ at $h\tau = \pi/4$) decreases when $L$ increases,
hence our state filtration protocol provides various target states with counterintuitive scaling of preparation time.
In the thermodynamic limit,
all the bright eigenvalues go to $0$,
and the target state is immediately created after the first measurement.
We note that this scenario is not exceptional;
rather, it occurs for multiple values of the control parameter $h\tau$,
due to the equally spaced energy spectrum.
} 
\label{fig:domBriEv}
\end{figure*}

\end{widetext}

\onecolumngrid
\clearpage

\begin{center}
    {\bf 
    {\large 
    {\em Supplementary Material:} \\ \smallskip
    Preparation of cat states in many-body eigenbasis via non-local measurement
    }
    }
\end{center}
\setcounter{equation}{0}
\setcounter{figure}{0}
\setcounter{table}{0}
\makeatletter
\renewcommand\thesection{SM Note \arabic{section}}
\renewcommand\thesubsection{\thesection.\arabic{subsection}}
\renewcommand{\theequation}{S\arabic{equation}}
\renewcommand{\thefigure}{S\arabic{figure}}
\setcounter{secnumdepth}{2}
\setcounter{tocdepth}{2}

\section{Electrostatic analogy}
\begin{figure}[ht]
\centering
\includegraphics[width=0.17\linewidth]{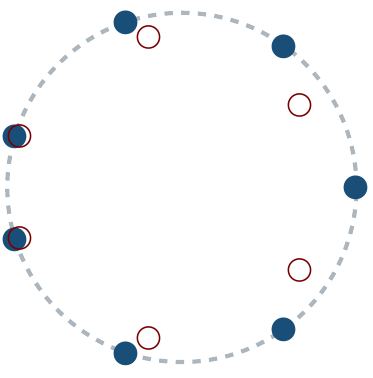}
\caption{The electrostatic analogy of bright eigenvalues $|\zeta_k|<1$. 
The blue circles are the charges $p_l$ located at $\exp(-i\tau E_l)$, and the red open circles are the equilibrium points of the composite force field,
or bright eigenvalues $\zeta_k$.
Here we use the spin-1 XY model with $L=6$, and Eq. \eqref{neel} as $\ket{\psi_{\rm r}}$. $h\tau = \pi/L -0.05$.} 
\label{fig:charge}
\end{figure}

Here we explain in detail the electrostatic analogy
for the bright eigenvalues ($|\zeta|<1$) of
the elementary filtration operator $\mathcal{F}^\tau_{\psi_{\rm r}}$ \cite{yin2019,Liu2022a}.
Starting from the expression of $\mathcal{F}^\tau_{\psi_{\rm r}}$
(below for simplicity we use ${\cal F}$)
in Eq. \eqref{eq:generalformula},
and applying the matrix determinant lemma to its characteristic polynomial,
we can obtain
\begin{equation}\label{matrixLemma}
\begin{aligned}
    0=&\text{det} \left[ \zeta \mathbbm{1} - {\cal F} \right]
    = \text{det} \left[ \zeta \mathbbm{1} - {U}(\tau) 
    + \ket{\psi_\text{r}}\bra{\psi_\text{r}} {U}(\tau) \right] \\
    =& {\zeta} \,\text{det} \left[ \zeta \mathbbm{1} - {U}(\tau) \right]
     \bra{\psi_\text{r}} \left[ \zeta \mathbbm{1} - {U}(\tau) \right]^{-1} \ket{\psi_\text{r}}.
\end{aligned}
\end{equation}
The bright eigenvalues are given by
$0 = \bra{\psi_\text{r}}
\left[ \zeta \mathbbm{1} - {U}(\tau) \right]^{-1} \ket{\psi_\text{r}}$,
which can be spectrally decomposed as
\begin{equation}\label{bright}
    \sum_{l=1}^w
    {\sum_{m=1}^{g_l} | \langle {\psi_\mathrm{r}}| {E_{lm}} \rangle |^2 
    \over
    e^{-iE_l\tau} - \zeta} = 0,
\end{equation}
where $w$ is the number of distinct energy phases $e^{-i\tau E_l}$,
$\ket{E_{lm}}$ ($m>1$) are the eigenstates of $U(\tau)$
that share the same eigenvalue $e^{-i\tau E_l}$,
and $g_l$ is the degeneracy of $e^{-i\tau E_l}$.
Thus, there will be $w-1$ roots of the summation,
since it is a polynomial of degree $w-1$.
Including the trivial $\zeta=0$ from Eq.~\eqref{matrixLemma},
there are $w$ bright eigenvalues/states.
Let us denote the overlap between the removal state
and all the degenerate energy eigenstates
by $p_l := \sum_m^{g_l} | \langle {\psi_\text{r}} | {E_{lm}} \rangle |^2$. 
With a new notation, Eq. \eqref{bright} becomes
\begin{equation}\label{forceF}
\begin{aligned}
	&\sum_l p_l \,{e^{iE_l\tau}-\zeta^\ast \over |e^{-iE_l\tau}-\zeta|^2} =0
    \Rightarrow \sum_l \,{\lambda_l \over 2\pi\varepsilon_0 \cdot r_l} \hat{r}_l =0.
\end{aligned}
\end{equation}
Hence, 
the last expression indicates a composite electric field,
stemming from infinite uniformly-charged wires,
with charge density $p_l\cdot 2\pi\varepsilon_0 \mapsto \lambda_l$ 
($\varepsilon_0$ is the vacuum permittivity).
These wires are pieced into the complex plane, 
located at $e^{-i\tau E_l}$ on the unit circle, 
with ${\zeta-e^{-iE_l\tau} \over |\zeta-e^{-iE_l\tau}|} \mapsto \hat{r}_l$ 
standing for the unit vector,
pointing from $\exp(-iE_l\tau)$ to $\zeta$
(we reverse the unimportant sign).
We note that there exists a constraint for $p_l$,
i.e. $\sum_{l=1}^w p_l =1$.
Therefore,
the eigenvalues $\{\zeta_k\}$ are the equilibrium points of the composite force field Eq. \eqref{forceF}.
Naturally, we expect that all the eigenvalues are inside the unit circle,
and specifically lie in the convex hull
constructed from the energy phases $e^{-iE_l\tau}$.
See the schematics of the charge picture in Fig. \ref{fig:charge}. 

The electrostatic analogy provides intuitive insight
into the distribution of bright eigenvalues $\zeta$.
A weak charge/overlap $p_l$ positions an eigenvalue $\zeta_k$ close to $e^{-i\tau E_l}$,
balancing weaker local forces against stronger distant ones.
Likewise, when two charges are close on the unit circle,
a local equilibrium forms nearby, largely unaffected by distant charges. 
In our problems, these intuitions identify the dominant bright eigenvalue—closest to the unit circle and slowest to decay—which governs the convergence of $Q_n$.
Using this electrostatic mapping,
we obtain the analytical results for the spin-1 XY model presented below.

\section{System size scaling, $\theta_0$ dependence of $n_\epsilon$}
The state filtration process relies on 
the depletion of all bright states,
hence the efficiency is set by their decay rates, 
determined by $|\zeta_k|$.  
Both the $|\zeta_k|$ and the number of bright states
are influenced by the system size \(L\) 
(see the electrostatic analogy).  
Thus, it is natural to ask how the filtration time \(n_\epsilon\) 
scales with \(L\).
We derive an asymptotic, exponential scaling for \(n_\epsilon(L)\). 

Filtration of $\ket*{\Psi_{\rm tar}^{(1)}}$:
When $h\tau_1=\pi/L$ and for arbitrary $\theta_0$,
the initial state $\ket{\psi_0(\theta_0)}$ has a nonzero overlap with the dominant bright state $\ket{\zeta_d}$
(the one with the largest $|\zeta_d|<1$), 
whose decay controls the filtration efficiency.
Consequently, we find that
\begin{equation}\label{scaling0}
    n_\epsilon \sim {2^{L} \over 8} 
    \ln \left[ {1+(-1)^{L} \cos(L\theta_0) \over 1-(-1)^{L} \cos(L\theta_0) } {1 \over \epsilon} \right].
\end{equation}
However, when $L\theta_0$ is an odd (even) multiple of $\pi$ for even (odd) $L$,
$|\langle \zeta_d | {\psi_0} \rangle|^2 \propto  1+(-1)^{L} \cos(L\theta_0) =0$,
and the dominant bright state does not contribute.
In that case the next-slowest decay bright state governs the scaling, giving
\begin{equation}\label{scaling1}
    n_\epsilon \sim {2^L \over 4L} \ln(L/\epsilon).
\end{equation}
Both of these predictions are in excellent agreement with our numerical results shown in Fig.~\ref{fig:QnGHZ} and the next section where a more general value of $\theta_0$ is considered.

Filtration of $\ket*{\Psi_{\rm tar}^{(2)}}$: 
Choosing $h\tau_2 = \pi/(L-1)$ for simplicity, we find
\begin{equation}\label{scaling3}
     n_\epsilon \sim {2^L \over 4(L+1)} 
     \ln 
     \left\{ 
            {1 \over 2L}   
            {1+L^2 - 2(-1)^L L \cos[(L-1)\theta_0] \over 1+(-1)^L \cos [(L-1)\theta_0]}  
            {1\over \epsilon}
    \right\}.
\end{equation}
In particular,
at special angles
$(L-1)\theta_0 = 2\pi k$ (even $L$) or $(2k+1)\pi$ (odd $L$),
$\langle \zeta_d | {\psi_0} \rangle$ is minimized,
and in the meantime  $\langle {\Psi_{\rm tar}^{(2)}} | \psi_0 \rangle$ reaches its maximum.
In this optimal case, Eq. \eqref{scaling3} attains its lower bound,
i.e. $n_\epsilon \sim {2^L \over 4(L+1)} \ln[(1-L)^2/4L\epsilon]$.
We present in Fig. \ref{fig:QnGHZ} the numerical results for this optimal choice of $\theta_0$ which
match well with the theory.

To derive Eqs. (\ref{scaling0}-\ref{scaling3}),
we employ the electrostatic analogy
to locate the dominant bright-state eigenvalue $\zeta_d$ for each choice of $h\tau$.
Applying the perturbation method to Eq. \eqref{bright}
yields the asymptotic expression of $|\zeta_d|$.
Then with the bright state expressions given by simple linear algebra \cite{Liu2022a},
the explicit decomposition of $\ket{\psi_n}$ (Eq. \eqref{eq:psi_n_decomposition})
can be obtained and used to solve $Q_n = 1-\epsilon$ (Eq. \eqref{Qn}).

\section{Filtration of $\ket*{\Psi_{\rm tar}^{(1)}}$ for general $\theta_0$}
\begin{figure}[h]
\centering
\includegraphics[width=0.7\linewidth]{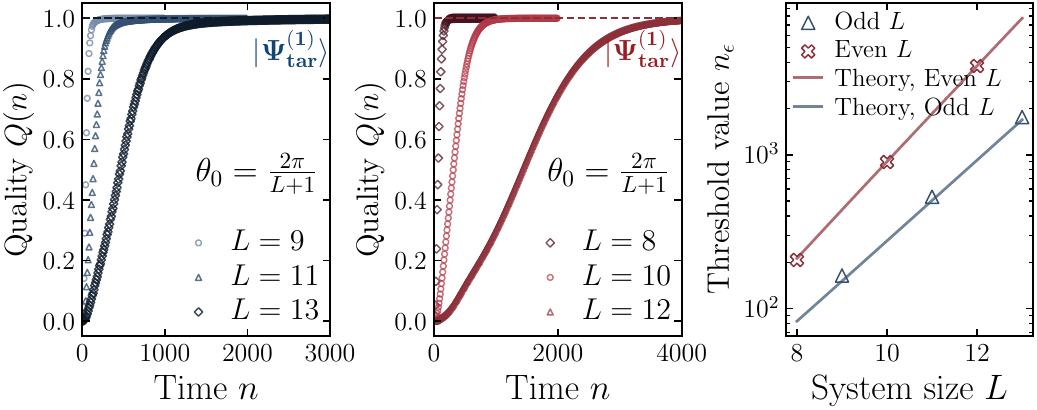}
\caption{
The filtration quality $Q_n$ for $\ket*{\Psi_{\rm tar}^{(1)}}$ (left, middle) 
as a function of the number of measurements $n$. 
The model here is the spin-1 XY chain (\ref{ham}) with $J=1, h=1$.
The removal state is the product state (Eq. (8) in the main text).
The initial angle is $\theta_0=2\pi/(L+1)$,
hence the parity of $L$ affects $n_\epsilon$
(see Eq. \eqref{scaling0}).
(right) The number of measurements, $n_\epsilon$, required to reach the fidelity $Q_n = 1-\epsilon$,
exhibits an exponential law of the system size $L$.
Here $\epsilon=0.01$.
The numerical results (markers) fit nicely with the theory
(Eq. \eqref{scaling0}, solid lines).
}
\label{fig:QnGHZgeneral}
\end{figure}

Here we show for a general choice of the initial angle $\theta_0$, the convergent behavior of the filtration quality $Q_n$.
Recall the analytical result for the filtration of $\ket*{\Psi_{\rm tar}^{(1)}}$ Eq. \eqref{scaling0},
and this equation holds unless the dominant bright state with the largest eigenvalue is orthogonal to the initial state dependent on $\theta_0$.
The parity of $L$ clearly affects the scaling with the system size $L$, which originates from the overlap $\langle \Psi_{\rm tar}^{(1)} | \psi_0 \rangle$ dependent on $L$'s parity.
As a complement to the main text showing the optimal case where the dominant bright state does not contribute,
we show here how Eq. \eqref{scaling0} match the numerics.
Choosing $\theta_0= 2\pi/(L+1)$, we plot in Fig. \ref{fig:QnGHZgeneral} the quality $Q_n$ as a function of time $n$,
as well as the threshold $n_\epsilon$ that guarantees $Q_n \ge 1-\epsilon$ when $n\ge n_\epsilon$.
Due to the influence of the parity of $L$, the threshold $n_\epsilon$ for even $L$ increases faster than the case of odd $L$, as predicted by our theory (solid lines, Eq. \eqref{scaling0}). 

\end{document}